\documentclass[
reprint,
superscriptaddress,
showpacs,
preprintnumbers,
nofootinbib,
amsmath,
amssymb, 
aps,
prd,
usenames,
dvipsnames,
floatfix
]{revtex4-2}

\pdfoutput=1
\usepackage{multirow}
\usepackage{booktabs}

\usepackage{xcolor}
\usepackage{slashed,multirow,relsize,soul,feynmp-auto,tikz}
\usepackage{mathrsfs} 
\usepackage{amsmath}
\usepackage{cancel}
\usepackage{bbold}
\usepackage{mathrsfs}
\usepackage{braket}
\usepackage{bm}
\usepackage{physics}
\usepackage{multirow,bigstrut}
\usepackage{xspace}
\usepackage{newunicodechar}
\newunicodechar{−}{-}
\usepackage{rotating}
\usepackage{ragged2e}
\usepackage{dcolumn}

\usepackage{fontawesome} 
\definecolor{blue-violet}{rgb}{0.33, 0.17, 0.89}

\usepackage{ifpdf}
\ifpdf        
  \usepackage{graphicx, hyperref, xcolor}     
\else     
  \usepackage[dvipdfmx]{graphicx, hyperref, xcolor}     
\fi
 \usepackage[capitalize]{cleveref}

\definecolor{rossoferrari}{HTML}{D9073D}
\definecolor{mediumblue}{HTML}{0000CD}
\definecolor{forestgreen}{HTML}{228B22}
\definecolor{desy_blue}{HTML}{009EE2}
\definecolor{desy_orange}{HTML}{FD8800}
\hypersetup{
setpagesize=false,
bookmarksnumbered=true,%
bookmarksopen=true,%
colorlinks=true,%
linkcolor=rossoferrari,
urlcolor=mediumblue,
citecolor=mediumblue,
linktocpage=false
}

\renewcommand{\abs}[1]{\left\lvert #1 \right\rvert}

\makeatletter
\@addtoreset{equation}{section}

\makeatother

\usepackage{tikz}
\usepackage{tikz-feynman}
\tikzfeynmanset{compat=1.0.0}

\newcommand{\gitlink}{\href{https://github.com/mhostert/tauALP}{{\large\color{blue-violet}\faGithub} \textsc{g}it\textsc{h}ub}\xspace}
\newcommand{\pythia}{\textsc{Pythia}\xspace}

\definecolor{wildstrawberry}{rgb}{1,0.263,0.643}

\newcommand{\beq}{\begin{equation}}
\newcommand{\eeq}{\end{equation}}
\newcommand{\be}{\begin{equation}}
\newcommand{\ee}{\end{equation}}
\allowdisplaybreaks

\begin{document}

\title{Long-lived Axion-Like Particles from Tau Decays}

\author{Yohei Ema}
\email{yohei.ema@cern.ch}
\affiliation{CERN Theory Division, CH-1211 Geneva 23, Switzerland}

\author{Patrick J.~Fox}
 \email{pjfox@fnal.gov}
\affiliation{Theory Division, Fermilab, Batavia, IL 60510, USA}

\author{Matheus Hostert}
\email{mhostert@g.harvard.edu}
\affiliation{Department of Physics \& Laboratory for Particle Physics and Cosmology, Harvard University, Cambridge, MA 02138, USA}

\author{Tony Menzo}
\email{menzoad@mail.uc.edu}
\affiliation{Department of Physics, University of Cincinnati, Cincinnati, Ohio 45221, USA}

\author{Maxim Pospelov}
\email{pospelov@umn.edu}
\affiliation{School of Physics and Astronomy, University of Minnesota, Minneapolis, MN 55455, USA}
\affiliation{William I. Fine Theoretical Physics Institute, School of Physics and Astronomy, University of Minnesota, Minneapolis, MN 55455, USA}

\author{Anupam Ray}
 \email{anupam.ray@berkeley.edu}
\affiliation{Department of Physics, University of California Berkeley, Berkeley, California 94720, USA}

\author{Jure Zupan}
\email{zupanje@ucmail.uc.edu}
\affiliation{Department of Physics, University of Cincinnati, Cincinnati, Ohio 45221, USA}

\date{\today}

\begin{abstract}
Axion-like particles (ALPs) are well-motivated examples of light, weakly coupled particles in theories beyond the Standard Model.
In this work, we study long-lived ALPs coupled exclusively to leptons in the mass range between $2 m_e$ and $m_\tau - m_e$.
For anarchic flavor structure the leptophilic ALP production in tau decays or from ALP-tau bremsstrahlung is enhanced thanks to derivative couplings of the ALP and can surpass production from electron and muon channels, especially for ALPs heavier than $m_\mu$.
Using past data from high-energy fixed-target experiments such as CHARM and BEBC we place new constraints on the ALP decay constant $f_a$, reaching scales as high as $\mathcal{O}(10^8)$~GeV in lepton-flavor-violating channels and $f_a \sim \mathcal{O}(10^2)$~GeV in lepton-flavor-conserving ones. 
We also present projections for the event-rate sensitivity of current and future detectors to ALPs produced at the Fermilab Main Injector, the CERN SPS, and in the forward direction of the LHC.
We show that SHiP will be sensitive to $f_a$ values that are over an order of magnitude above the existing constraints.
\end{abstract}

\preprint{CERN-TH-2025-123, FERMILAB-PUB-25-0408-T, N3AS-25-011}

\maketitle

\twocolumngrid
 
\section{Introduction}
\label{sec:LLP}

The search for light, very weakly coupled particles has emerged as a crucial avenue for exploring physics beyond the Standard Model (SM). This approach is complementary to the more traditional strategy of searching for new heavy particles at increasingly higher energies. Over the last two decades, significant efforts have been devoted to defining a ``space of possibilities'' for feebly interacting particles, commonly referred to as the ``dark sector physics'' (see, {\em e.g.},~\cite{Beacham:2019nyx,Battaglieri:2017aum,Lanfranchi:2020crw} and references therein).

One of the better motivated examples of such physics is a light scalar, $a$, derivatively coupled to the SM fermions $\psi^i_{L,R}$, 
\beq 
\label{eq:Lint:general}
{\cal L}_{\rm int}\propto \big( \bar \psi^i_{L,R} \gamma^\mu \psi^j_{L,R}\big) \times \partial_\mu a/f_a. 
\eeq
These interactions are of canonical dimension 5, and are thus suppressed by an energy scale, conventionally denoted as $f_a$. Similar types of derivatively coupled operators can be built from the SM gauge fields, 
\beq
\big(\epsilon^{\mu\nu\alpha\beta}\Tr [A_\nu\partial_\alpha A_\beta] \big) \times \partial_\mu a /f_a,
\eeq
with $A_\mu=A_\mu^aT^a$ denoting any of the $SU(3)_c\times SU(2)_L\times U(1)_Y$ gauge fields. 
A notable feature of these interactions is the absence of large perturbative corrections to the $a$ mass, $m_a$, from loops involving the SM fermions $\psi_{L,R}^i$. At a nonperturbative level, however, the interactions with gluons do generate a mass term for $a$, which plays a crucial role in the QCD axion mechanism -- allowing the originally massless axion field to dynamically restore  CP conservation in QCD~\cite{Peccei:1977hh,Weinberg:1977ma,Wilczek:1977pj}. 

In general, a light scalar $a$ need not acquire its mass solely through the QCD anomaly. This more general possibility, where $m_a$ is treated as a free parameter, will be referred to as an axion-like particle (ALP). It is also worth noting that ALP interactions with the SM fermions constitute only a subset of a broader class of operators, which include also non-derivative and/or scalar couplings, as recently discussed in~\cite{Balkin:2024qtf,Delaunay:2025lhl}.

In the general interaction of \cref{eq:Lint:general}, the indices $i,j$ label different types of SM fermion, such as quarks and leptons and include different generations (respecting, of course, an overall gauge invariance of the current). Clearly, a very large variety of such couplings is possible, and without the knowledge of ultraviolet completion of these interactions no definitive predictions regarding their relative size can be made. In this paper, we will stay at the level of the effective field theories (EFTs), and will explore couplings of $a$ to leptonic currents. Our primary emphasis will be on the relatively unexplored class of interactions involving tau leptons.
The lowest dimension EFT Lagrangian is thus
\beq
\mathcal{L}_a=\mathcal{L}_\text{kin}+\mathcal{L}_{\rm int},
\eeq
where the canonical kinetic term for $a$, 
\beq
\mathcal{L}_\text{kin}=\frac12(\partial_\mu a)^2 - \frac12m_a^2a^2,
\eeq
is supplemented by the interaction terms involving just the lepton currents,
\begin{align}
\label{eq:La}
	\mathcal{L}_{\rm int} &= \frac{\partial_\mu a}{2 f_a}
	\sum_{i,j} \left[g_L^{ij} \bar{L}_{i} \gamma^\mu P_L L_{j}
	+ g_R^{ij} \bar{\ell}_{i} \gamma^\mu P_R \ell_{j}\right],
\end{align}
where we made the chirality explicit.
In this expression, $f_a$ is the overall suppression scale, while the  indices on the dimensionless real coefficients $g_L^{ij}$ and $g_R^{ij}$ run over the three generations of SM leptons, $i,j=1,2,3$.

In the analysis and numerical calculation below we limit ourselves to the situation where $g_L^{ij}$ and $g_R^{ij}$ have the same magnitude but opposite sign, {\em i.e.}, $g_{\ell_1 \ell_2} = g_L^{\ell_1 \ell_2} = -g_R^{\ell_1\ell_2}$,  and use the following short-hand notation, 
\beq
\frac{1}{f_{\ell_1 \ell_2}} = \frac{g_{\ell_1\ell_2}}{f_a}.
\eeq

The Lagrangian in~\cref{eq:La} has two phenomenologically important features. 
The off-diagonal couplings  $f_{e\tau}^{-1}$ and $f_{\mu\tau}^{-1}$ induce flavor-changing neutral current (FCNC) decays $\tau\to ea$ and $\tau \to \mu a$. Since these are induced by dimension-5 operators they are parametrically enhanced by $(m_W^2/m_\tau f_{\ell \tau})^2$ compared to the SM decay rates, which are suppressed by Fermi coupling constant $G_F\propto 1/m_W^2$. Therefore, nonzero $f_{\mu\tau}^{-1}$ and $f_{e\tau}^{-1}$ can efficiently produce $a$ in any process that results in $\tau$ leptons. 

The second notable feature of the Lagrangian in~\cref{eq:La} is the additional suppression of the partial decay widths $a \to \mu^+\mu^-$ and, in particular, $a \to e^+e^-$. The currents that the axion couples to are conserved in the limit of massless leptons, so the two decays widths are proportional to lepton masses.  In particular, $\Gamma(a \to \mu^+\mu^-)\propto m_a (m_\mu/f_{\mu\mu})^{2}$ and $\Gamma(a \to e^+e^-)\propto m_a(m_e/f_{ee})^{2}$, and are thus suppressed by the relative smallness of the leptonic masses $m_\mu$ and $m_e$ relative to the fiducial mass scale set by $m_a$, which can be as large as $\sim 1$\,GeV. As a result, $a$ is expected to be relatively long-lived, potentially giving rise to displaced decay signatures --- a possibility that we explore in this work.

\begin{figure}[t!]
    \centering   
    \includegraphics[width=0.42\textwidth]{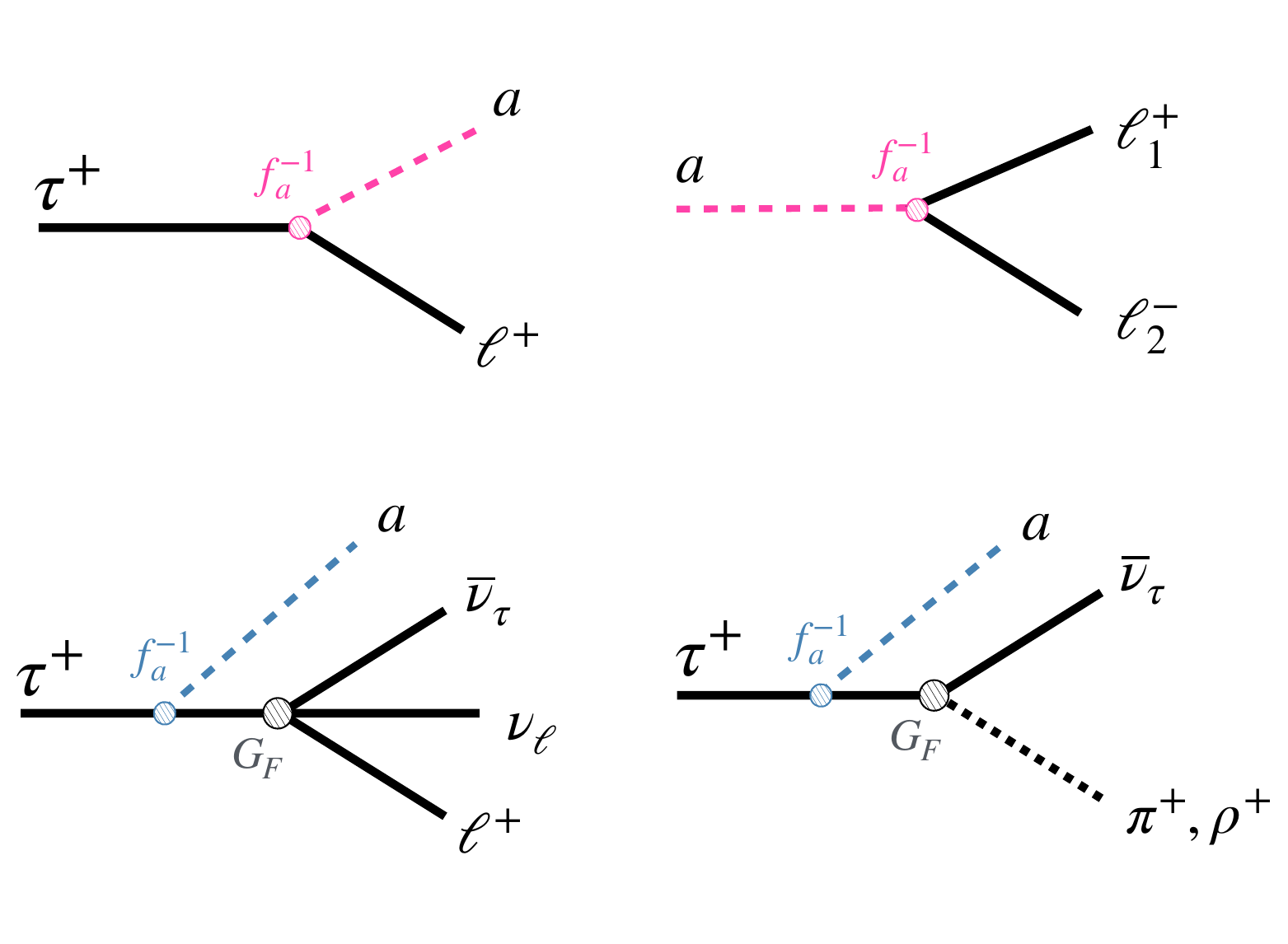}
    \caption{
    Tau decays involving LFV (top left) and LFC (bottom row) ALPs, where, in addition, the
    ALP can also attach to $\bar{\nu}_\tau$ (not shown).
    The ALP decay is shown in the top right.
    \label{fig:diagrams}
    }
\end{figure}

Even though we restricted the parameter space of possible ALP couplings to just the ones coupling ALPs to leptons, there is still considerable freedom in the choices for these couplings. At the qualitative level, though, the bulk of the phenomenology can be captured by considering the following lepton-flavor-violating (LFV) and lepton-flavor-conserving (LFC) benchmark scenarios:
\begin{enumerate}
    \item {\em Anarchical LFV.} This scenario assumes a ``democratic" pattern of couplings, without any particular suppression of the off-diagonal couplings:
    \begin{equation}\label{eq:flavor_anarchy}
        g_{\tau\tau}\sim g_{\mu\tau}\sim g_{\mu\mu}\sim g_{ee},~\text{etc.}
    \end{equation}
    The main source of $a$ will be the FCNC production from $\tau \to \ell a$, while the dominant $a$ decay channel will be into the heaviest kinematically available lepton pair. 

    \item{\em Hierarchical LFV.} In this scenario the flavor-changing couplings are assumed to be smaller,
\begin{equation}\label{eq:flavor_hierarchy}
   g_{\tau\tau}  \sim g_{\mu\mu}\sim g_{ee} > g_{\mu\tau},\, g_{e\tau}, g_{e\mu}~, 
\end{equation}
    but not negligible, so that the FCNC decays $\tau \to \ell a$ still dominate ALP production. 
    Within this scenario, we will be able to explore how the size of the hierarchy between LFV and LFC couplings impacts the experimental reach.

    \item{\em Universal LFC.} In this benchmark, the FCNC couplings are set to zero, while the diagonal couplings have no particular pattern
    \begin{equation}\label{eq:flavor_conserving_universal}
         g_{\tau\tau}  \sim g_{\mu\mu}\sim g_{ee};~~ g_{\mu\tau}= g_{e\tau}=g_{e\mu}=0.
    \end{equation}
    In this scenario, the ``bremsstrahlung" production of $a$ in tau decays ({\em e.g.} $\tau\ \to \pi\nu a;~\pi\pi\nu a$ etc) will be important, and the longevity of $a$ will be achieved only for $m_a < 2m_\mu$.

    \item{\em Non-universal LFC.} In this benchmark, ALPs only have flavor diagonal couplings to leptons, and these exhibit a hierarchical pattern. We are particularly interested in the case, where ALPs produced in tau decays are always long-lived, which occurs for a flavor non-universal coupling choice of 
    \begin{equation}\label{eq:flavor_conserving_nonuniversal}
         g_{\mu\mu}  \ll  g_{\tau \tau};~~ g_{\mu\tau}= g_{e\tau}=g_{e\mu}=0.
    \end{equation}
    Depending on whether $g_{ee}$ is sizeable or not, the ALP will decay mostly to $e^+e^-$ or to $\gamma\gamma$, but it will be typically long-lived in the mass range of interest.
    We consider both cases: a {\em $\mu$-phobic ALP}, with    
    \beq
    \label{eq:e-tau}
    g_{\mu \mu} \ll g_{ee} \sim g_{\tau\tau},
    \eeq
    and a {\em $\tau$-philic scenario} with 
    \beq
    \label{eq:tau:philic}
    g_{\mu \mu}, g_{ee} \ll g_{\tau\tau}.
    \eeq 
\end{enumerate}

Our goal is to obtain existing constraints on the above scenarios and provide projections both for the upgrades of the existing experiments, as well as for future  projects such as SHiP, focusing on processes involving tau leptons. 
The main finding is that beam dump and collider searches for displaced ALPs from tau decays provide better sensitivity to such ALPs than the precision studies of tau decays.
The displaced ALP decays to $\mu^+\mu^-$ and muon-electron pairs are a currently under-explored signature of  LFV and LFC models of ALPs coupled exclusively to lepton currents, even though both channels are generically available and even dominant.  
We make projections for these types of searches for SHiP and the next generation of the FASER experiment. In the LFC case, we also derive novel constraints on light ALPs with displaced decays to $e^+e^-$ pairs. 

The paper is organized as follows. In section~\ref{sec:ALPprod} we derive the branching ratios for production  of ALPs in $\tau$ decays, as well as ALP decay rates. In section~\ref{sec:signatures} we discuss the resulting ALP production rates at past, existing, and future experiments. 
Our main results, the current and projected limits on the ALP parameter space, are given in section~\ref{sec:elimits}.  Section~\ref{sec:conclusions} contains our conclusions, while further details on the calculation of ALP production and decay rates are relegated to appendix \ref{app:details}.

\section{ALP production and decay} 
\label{sec:ALPprod}

In this section, we summarize the ALP production and decay rates.
We focus on the production of ALPs in the decays of tau leptons, see \cref{fig:diagrams}.
The corresponding differential decay rate distributions, in $\tau$ rest frame, are shown in \cref{fig:diff_branchings}. 
Additional details are provided in appendix~\ref{app:details}.
The main production mechanism of $\tau$ leptons in beam dump experiments are the decays of $D_s$ meson, since the branching ratio $\mathrm{Br}(D_s^\pm\to \tau^\pm\overset{(-)}{\nu_\tau})\simeq 5.36\%$ is quite large. 
Other sub-leading production channels include $\mathrm{Br}(D^\pm\to \tau^\pm\overset{(-)}{\nu_\tau})= 1.20 \times10^{-3}$ and $
\mathrm{Br}(\psi(2S) \to \tau^+\tau^-)= 3.10 \times10^{-3}$ decays~\cite{ParticleDataGroup:2024cfk}.
The dominant production of ALPs is then either from 2-body decays of tau leptons (for the LFV scenarios) or from multibody decays of taus (for the LFC scenarios). 

\subsection{ALP production: LFV scenarios}
\label{sec:ALP:prod:LFV}

If the LFV couplings of the ALP are appreciable, the ALP is dominantly produced through the 
flavor-changing decay of tau leptons, $\tau \to \mu a, e a$. This is the case for both the anarchical and hierarchical LFV scenarios,  \cref{eq:flavor_anarchy,eq:flavor_hierarchy}, respectively.
The decay rate is given by
\begin{align}
	\Gamma(\tau \to \ell a) &= \frac{\vert g_{\ell\tau}\vert^2 m_\tau^3}{64\pi f_a^2}
	\left(1 + \frac{m_\ell}{m_\tau}\right)^2
	\left[\left(1 - \frac{m_\ell}{m_\tau}\right)^2 - \frac{m_a^2}{m_\tau^2}\right]  \nonumber \\ 
	&\times
	v(m_\tau, m_\ell, m_a),
\end{align}
where $\ell = \mu, e$, and $v$ is the separation velocity,
\begin{align}
	v(M, m_1, m_2) = \sqrt{1-\frac{2(m_1^2 + m_2^2)}{M^2}+\frac{(m_1^2 - m_2^2)^2}{M^4}}.
\end{align}
For instance, for $m_a$ = 0.1 GeV, we find
\begin{equation}
\label{eq:LFV:numerical:example}
	\mathrm{Br}(\tau \to \ell a) = 1.2\times 10^{-4} \vert g_{\ell\tau}\vert^2 \left(\frac{10^7\,\rm{GeV}}{f_a}\right)^2,
\end{equation}
with $\ell=e,\mu$.
Note that the branching ratio for ALP production is enhanced by the small SM tau decay width, i.e., it is enhanced by $G_F^{-2}$, where  $G_F$ is the Fermi constant.
Consequently, $\mathrm{Br}(\tau \to \ell a)$ is sizable even for relatively large values of $f_a$, as long as $g_{\ell \tau}$ is not too small. A large value of $f_a$, on the other hand, implies a long-lived ALP.

Here we note that, if $m_a < m_\mu - m_e$, the $\mu \to e a$ decays are kinematically allowed. In this ALP mass range the searches for $\mu \to e a$ decays place stringent constraints on ALP LFV scenarios as long as $g_{e \mu}$ coupling is comparable to the $g_{ \ell \tau}$ couplings~\cite{Jodidio:1986mz,TWIST:2014ymv,PIENU:2020loi,Calibbi:2020jvd}.

\begin{figure}[t!]
    \centering   \includegraphics[width=0.49\textwidth]{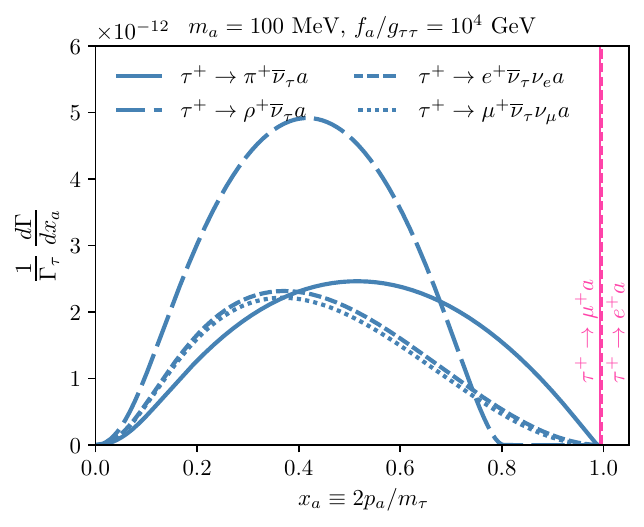}
    \caption{
    Differential branching ratios for various tau decays involving ALPs, as indicated, for a sample ALP mass $m_a = 100$~MeV, and ALP decay constant $f_a=10^4$\,GeV, under the $\tau$-philic scenario hypothesis of \cref{eq:tau:philic}.
    Here,  $x_a = 2 p_a / m_\tau$ is the fractional momentum carried by the ALP in the tau rest frame ($p_a$ is the magnitude of the ALP momentum).
    We also indicate with vertical purple lines the value of $x_a$ corresponding to the two mono-energetic 2-body decay modes, $\tau\to \ell a$.
    \label{fig:diff_branchings}
    }
\end{figure}

\begin{figure}[t]
    \centering
    \includegraphics[width=0.49\textwidth]{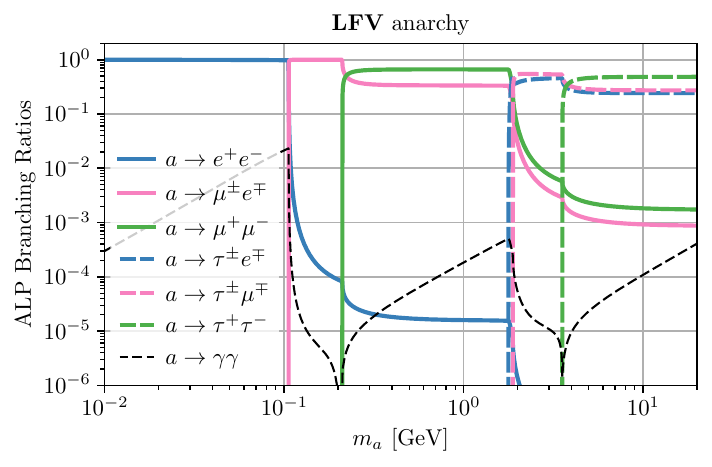}
    \includegraphics[width=0.49\textwidth]{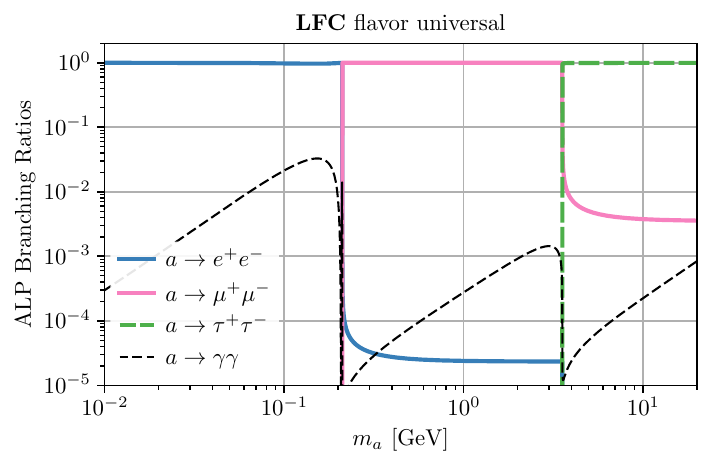}
    \includegraphics[width=0.49\textwidth]{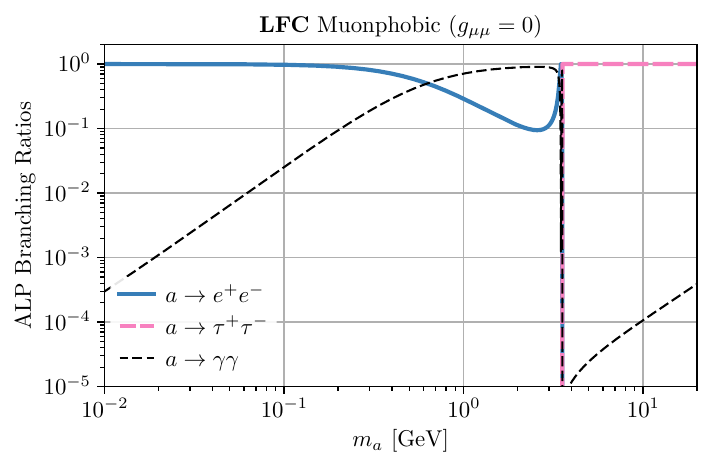}
    \caption{The ALP decay branching ratios 
    in the LFV anarchy (top), LFC flavor universal (middle), and LFC $e-\tau$ coupled (bottom) scenarios.
     }
    \label{fig:alpBRs}
\end{figure}

\subsection{ALP production: LFC scenarios}
\label{sec:ALP:prod:LFC}

If flavor-changing couplings $g_{\ell \tau}$ are vanishingly small, \emph{i.e.}, for the LFC scenarios in \cref{eq:flavor_conserving_universal,eq:flavor_conserving_nonuniversal}, 
tau leptons can still produce ALPs. The ALP production then proceeds via ALP bremsstrahlung modifying any of the SM $\tau$ decay channels, see the bottom row in \cref{fig:diagrams}.
For simplicity, we focus on the decay modes 
$\tau^- \to \pi^- \nu_\tau$, $\tau^- \to \pi^- \pi^0 \nu_\tau$, and $\tau^- \to \ell^- \bar{\nu}_\ell \nu_\tau$,
which account for $\sim 70\,\%$ of the SM $\tau$ decay width. 
Including more decay modes would result in a marginally tighter bound on (better sensitivity to) the ALP couplings.

To calculate the ALP production via bremsstrahlung in $\tau \to \pi \nu_\tau a$, we use the following effective Lagrangian for the SM rate
\begin{align}
	\mathcal{L}_\mathrm{eff} = -\frac{G_F}{\sqrt{2}} V_{ud}^*f_\pi (\partial_\mu \pi^-)\bar{\tau}\gamma^\mu
	(1-\gamma_5)\nu_\tau + (\mathrm{h.c.}),
\end{align}
where $V_{ud} \simeq 0.97$ is the CKM matrix element and $f_\pi = 130\,\mathrm{MeV}$ the pion decay constant.
The differential $\tau \to \pi \nu_\tau a$ decay rate is given by\footnote{Note that since ALP couplings in \cref{eq:La}  are SU$(2)_L$-invariant, there are no so-called ``weak-violating" ALP type contributions to these decays~\cite{Altmannshofer:2022ckw,Jiang:2024cqj}.}
\begin{align}
	\frac{d\Gamma(\tau \to \pi \nu_\tau a)}{dE_a}
	&= \frac{\abs{g_{\tau\tau}}^2 G_F^2 \abs{V_{ud}}^2 f_\pi^2 m_\tau^2(\bar{m}_\tau^2 - m_\pi^2)^2}{256\pi^3  f_a^2 \bar{m}_\tau^2}
	\nonumber \\
	&\times \left[1 - \frac{m_a^2(m_\tau^2 + \bar{m}_\tau^2)}{(m_\tau^2 - \bar{m}_\tau^2)^2}\right]
	v(m_\tau,\bar{m}_\tau,m_a),
	\label{eq:dGamma_tau2pinua}
\end{align}
where $E_a$ is the ALP energy in the $\tau$-rest frame. Because of ALP bremsstrahlung from the $\tau$ leg, the intermediate $\tau$ line is off-shell, with the 
off-shell invariant mass squared $\bar{m}_\tau^2$ given by
\begin{align}
	\bar{m}_\tau^2 = m_\tau^2 + m_a^2 - 2m_\tau E_a.
\end{align}

Integrating the differential rate in \cref{eq:dGamma_tau2pinua} over $E_a\in[ m_a, (m_\tau^2 + m_a^2 - m_\pi^2)/2m_\tau]$,
gives $\Gamma(\tau \to \pi \nu_\tau a)$.
For example, for $m_a = 0.1\,\mathrm{GeV}$, we find 
\begin{align}
\label{eq:numerics:tau:pi:nutau:a}
	\mathrm{Br}(\tau \to \pi \nu_{\tau} a) = 1.5\times 10^{-12} \abs{g_{\tau\tau}}^2 \left(\frac{10^4\,\rm{GeV}}{f_a}\right)^2.
\end{align}
Compared to the $\tau \to \ell a$ decay rate in \cref{eq:LFV:numerical:example} this branching ratio is significantly more suppressed, if all couplings of ALP to leptons are ${\mathcal O}(1)$. The reason is that the $\tau \to \pi \nu_{\tau} a$ decay rate suffers double suppression, the factor of $G_F^2$ due to SM weak interaction, and $(g_{\tau\tau}/f_a)^2$ for the emission of the ALP. 
For LFC scenarios the phenomenologically relevant values of $f_a$ will be relatively low, $f_a \lesssim 10^4\,\mathrm{GeV}$.
This is true for the other tau decay channels involving ALPs, that we discuss below. 
For such low values of $f_a$ the ALP decays are displaced only for $m_a < 2 m_\mu$ or $\abs{g_{\mu\mu}} \ll 1$,  so that the dominant ALP decay is $a\to e^+e^-$, whose rate is suppressed by $m_e^2$ in the assumed LFC flavor patterns.

The decay mode $\tau^- \to \pi^- \pi^0 \nu_\tau$ is dominated by the intermediate $\rho$ meson decaying via its main decay mode, $\rho^- \to \pi^- \pi^0$.
Ignoring the $\rho$ width effects we can therefore  treat this 
ALP production channel as though it is due to a quasi three-body decay $\tau^- \to \rho^- \nu_\tau a$.
The effective Lagrangian relevant for the $\rho$ meson production can be written as~\cite{Donoghue:2022wrw}
\begin{align}
	\mathcal{L}_\mathrm{eff} = -\frac{G_F}{\sqrt{2}}V_{ud}^* m_\rho f_\rho \rho_\mu
	\bar{l}\gamma^\mu(1-\gamma_5) \nu_l + (\mathrm{h.c.}),
\end{align}
where $m_\rho \simeq 775\,\mathrm{MeV}$ is the $\rho$ meson mass and  $f_\rho = 216\,\mathrm{MeV}$  the $\rho$ decay constant. 
The differential decay rate is given by
\begin{align}
	&\frac{d\Gamma(\tau \to \rho \nu_\tau a)}{dE_a}
	= \frac{\abs{g_{\tau\tau}}^2 G_F^2 \abs{V_{ud}}^2 f_\rho^2 m_\tau^2\left(\bar{m}_\tau^2+ 2m_\rho^2\right)}{256\pi^3 f_a^2}
	\nonumber \\
	&~~~~~\times
	\left[1 - \frac{m_\rho^2}{\bar{m}_\tau^2}\right]^2
	\left[1 - \frac{m_a^2(m_\tau^2 + \bar{m}_\tau^2)}{(m_\tau^2 - \bar{m}_\tau^2)^2}\right]
	v(m_\tau,\bar{m}_\tau,m_a),
	\label{eq:dGamma_tau2rhonua}
\end{align}
Integrating the above differential rate from $E_a = m_a$ to $E_a = (m_\tau^2 + m_a^2 - m_\rho^2)/2m_\tau$ gives the $\Gamma(\tau \to \rho \nu_\tau a)$ decay rate.
For example, for $m_a = 0.1\,\mathrm{GeV}$, we find 
\begin{align}
	\mathrm{Br}(\tau \to \rho \nu_{\tau} a) = 2.3\times 10^{-12} \abs{g_{\tau\tau}}^2\left(\frac{10^4\,\rm{GeV}}{f_a}\right)^2,
\end{align}
and is thus numerically comparable to $\mathrm{Br}(\tau \to \pi \nu_{\tau} a)$,  \cref{eq:numerics:tau:pi:nutau:a}, as long as ALP is not too heavy.

\begin{figure*}[t]
    \centering
    \includegraphics[width=0.49\textwidth]{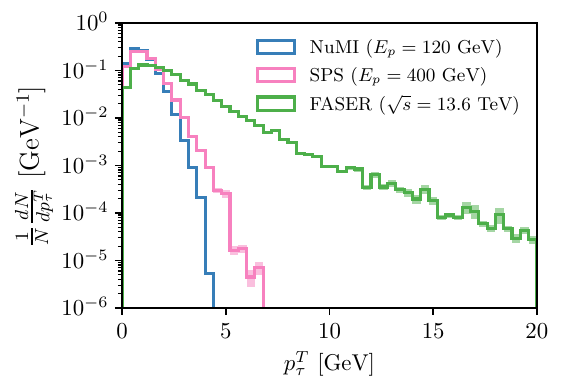}
    \includegraphics[width=0.49\textwidth]{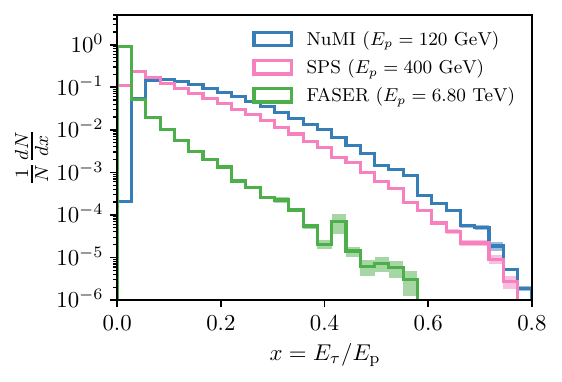}
    \caption{Left:
    the normalized differential tau production rate at the collision point as function of the tau transverse momentum $p_\tau^T$ (defined with respect to the $z$-axis, i.e., the axis of the proton beam)  for the three beams considered in this paper: NuMI, SPS, and FASER (as indicated).
    Right: the distribution of tau events as a function of the ratio between the total tau energy and the proton beam energy for the same three beams. 
    The shaded bands show the statistical uncertainties on Monte Carlo samples.  
    \label{fig:tau_spectra}
    }
\end{figure*}

Finally, the leptonic decay mode $\tau^- \to \ell^- \bar{\nu}_\ell \nu_\tau$ is governed by the four-Fermi interaction
\begin{align}
	\mathcal{L} = -\frac{G_F}{\sqrt{2}}\left[\bar{\nu}_\tau \gamma_\alpha (1-\gamma_5)\tau\right]
	\left[\bar{\ell}\gamma^\alpha (1-\gamma_5) \nu_{\ell}\right],
\end{align}
where $\ell = e, \mu$. The $\tau^-\to \ell^- \bar{\nu}_\ell \nu_\tau a$ channel is a four-body decay.
We perform the required phase space integrals analytically in appendix~\ref{app:details}, leading to the
following expression for the double differential decay rate
\beq
\begin{split}
	&\frac{d\Gamma(\tau\to\ell\bar{\nu}_\ell\nu_\tau a)}{dE_a  d \bar{m}_{\nu\nu}^2} = \frac{G_F^2 \abs{g_{\tau\tau}}^2m_\tau^2 \bar{m}_\tau^2}{1536 \pi^5 f_a^2} I(\bar{m}_\tau,\bar{m}_{\nu\nu})
 \\
	&\qquad \quad\times \left[1 - \frac{m_a^2 (m_\tau^2 + \bar{m}_\tau^2)}{(m_\tau^2 - \bar{m}_\tau^2)^2}\right]
	v(m_\tau,\bar{m}_\tau,m_a),
	\label{eq:dGamma_tau2lnunua}
\end{split}	
\eeq
where $E_a \in[ m_a, (m_\tau^2 + m_a^2 - m_\ell^2)/2m_\tau]$ is the ALP energy, $\bar{m}_{\nu\nu}^2\in [0,(\bar{m}_\tau - m_l)^2]$ is the invariant mass squared of the $\bar{\nu}_\ell \nu_\tau$ pair, and
\begin{align}
	I(\bar{m}_\tau, \bar{m}_{\nu\nu}) &= 
	\left[1+\frac{\bar{m}_{\nu\nu}^2 - 2m_l^2}{\bar{m}_\tau^2} + \frac{m_l^4 + m_l^2 \bar{m}_{\nu\nu}^2 - 2\bar{m}_{\nu\nu}^4}{\bar{m}_\tau^4}\right]
	\nonumber \\
	&\times 
	v(\bar{m}_\tau,\bar{m}_{\nu\nu},m_\ell).
	\label{eq:I_def}
\end{align}
In \cref{eq:dGamma_tau2lnunua} we only included ALP emissions from $\tau$ and $\nu_\tau$ lines since the one from $\ell$ and $\bar{\nu}_\ell$ are suppressed by $m_\ell/m_\tau$ at the level of the amplitude, and are thus never important for any of the LFC flavor structure we consider.

Integrating the differential rate in \cref{eq:dGamma_tau2lnunua} over $E_a$ and $\bar m_{\nu\nu}^2$ gives the $\Gamma(\tau\to\ell\bar{\nu}_\ell\nu_\tau a)$ decay rate.
For instance, for $m_a = 0.1\,\mathrm{GeV}$, we obtain for the two leptonic branching ratios,
\begin{align}
	\mathrm{Br}(\tau \to e\bar{\nu}_e\nu_\tau a) &= 1.2\times 10^{-12} \abs{g_{\tau\tau}}^2\left(\frac{10^4\,\rm{GeV}}{f_a}\right)^2~, \\
    \mathrm{Br}(\tau \to \mu\bar{\nu}_\mu\nu_\tau a) &= 1.1\times 10^{-12} \abs{g_{\tau\tau}}^2\left(\frac{10^4\,\rm{GeV}}{f_a}\right)^2.
    \label{eq:Br:tau:mu:numu:nutau:a}
\end{align}

\subsection{Other production modes}
\label{sec:other_production}

For universal LFC scenario, where $g_{\mu\mu}\sim g_{\tau\tau}$ one could consider including ALP production directly from $D_s, D^0$ and $D^+$ decays. 
For $D_s$ the relevant channel is $D_s \to \mu \nu_\mu a$. 
As a numerical example let us take $m_a = 0.1\,\mathrm{GeV}$, $f_a = 10^4\,\mathrm{GeV}$, giving $\mathrm{Br}(D_s \to \mu \nu_\mu a) \simeq 2.0\times 10^{-13}\vert g_{\mu\mu}\vert^2$, which is comparable to  $\mathrm{Br}(D_s^+ \to \tau^+ \nu_\tau) \mathrm{Br}(\tau \to \mu\bar{\nu}_\mu\nu_\tau a) \simeq  2.1 \times 10^{-13}\vert g_{\tau\tau}\vert^2$ (using \cref{eq:Br:tau:mu:numu:nutau:a} and $\mathrm{Br}(D_s^+\to \tau^+\nu_\tau)\simeq 5.4\%$).
For direct production of ALPs from $D^\pm$ and $D^0$ decays, the relevant channels are the exotic semileptonic decays $D\to K^{(*)} \mu\nu a$.
For instance, $\mathrm{Br}(D^+ \to \bar{K}^0 \mu^+ \nu_\mu a ) = \mathcal{O}(10^{-14}) \times \vert g_{\mu\mu}\vert^2$ for $f_a = 10^{4}\,\mathrm{GeV}$ and $m_a = 0.1\,\mathrm{GeV}$. 
Since $\sim 10\,\%$ of the charm quarks fragment into $D_s$ and the rest to $D^\pm$ and $D^0$ (sometimes via decays of excited charm mesons), the ALP production rate from these types of  decays is again comparable to the one from $\tau$-decays.

As we will demonstrate later, an LFC ALP with $m_a > 2 m_\mu$ is too short lived when $g_{\mu\mu} \sim g_{\tau\tau}$, making the limits and sensitivities discussed here largely irrelevant in this flavor-universal scenario. 
Below the dimuon threshold, the small-coupling parameter space is largely excluded anyway, so we neglect this contribution altogether, which is conservative.
Note that we also ignore the direct production of ALPs from $D_s \to \tau \nu_\tau a$ decays since this channel is severely phase-space suppressed. 

Since in the LFC scenarios the ALP production in $\tau$ decays is suppressed by both the Fermi constant and axion coupling, 
$\Gamma_{\tau \to X a} \propto O(G_F^2f_a^{-2})$, one may wonder if loop-induced processes may lead to enhanced production of ALPs in the FCNC decays of $B$ and $K$ mesons. If, for example, there is a direct coupling between an ALP and the $W$-boson pseudoscalar density $W_{\mu\nu}^a\widetilde W^{a\mu\nu}$ \cite{Izaguirre:2016dfi}, or if such a coupling is induced by the triangular lepton loop diagrams \cite{Dror:2017ehi,Dror:2017nsg}, there can be an enhanced production of ALPs due to one- or two-loop induced $(\partial_\mu a) \bar d_L \gamma_\mu s_L $ and $(\partial_\mu a) \bar s_L \gamma_\mu b_L $ operators. Such loop-suppressed production modes, {\em e.g.} $\Gamma_{K\to \pi a}$, can bypass the $G_F^2$ suppression and occur at the ${\mathcal O}(f_a^{-2})$ order. However, this type of production and ensuing constraints are model-dependent --- their evaluation requires a more detailed knowledge about the UV completion of the effective ALP interactions that we consider. Therefore, we concentrate on the tree-level production of ALPs, which are less susceptible to the unspecified model-building choices. 

\begin{figure*}[t]
    \centering
    \includegraphics[width=0.32\textwidth]{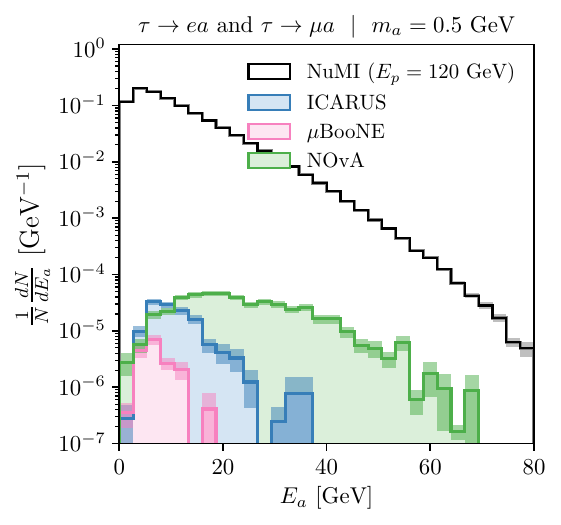}
    \includegraphics[width=0.32\textwidth]{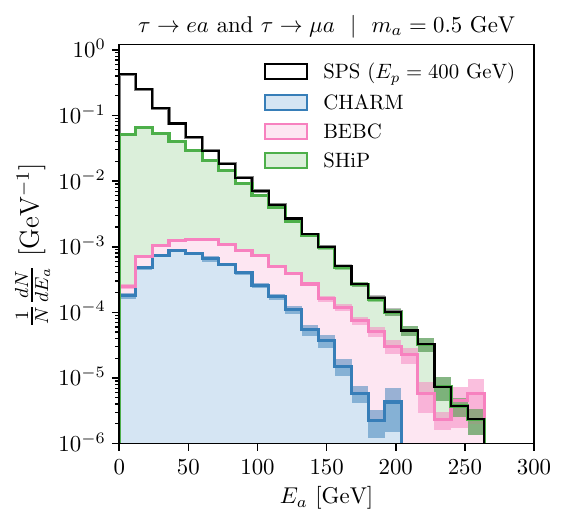}
    \includegraphics[width=0.32\textwidth]{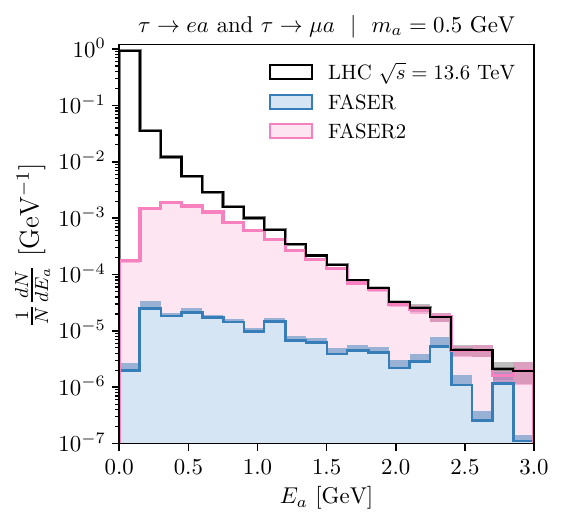}
    \includegraphics[width=0.32\textwidth]{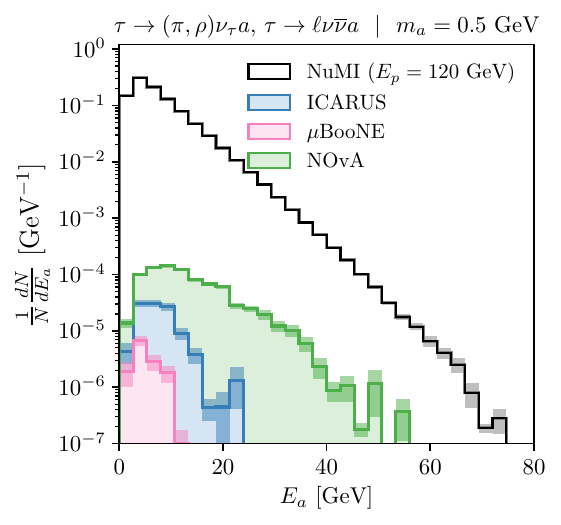}
    \includegraphics[width=0.32\textwidth]{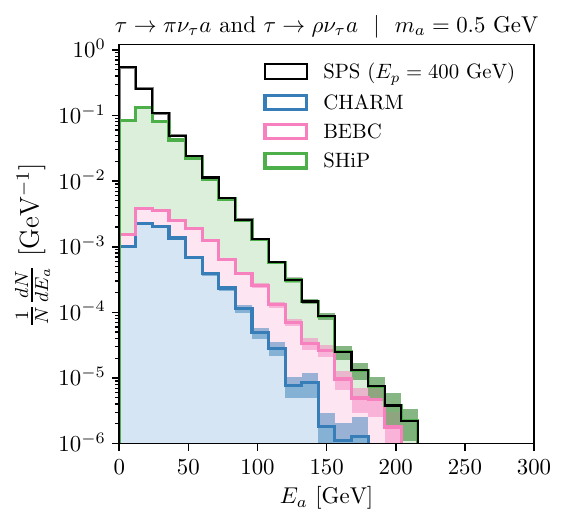}
    \includegraphics[width=0.32\textwidth]{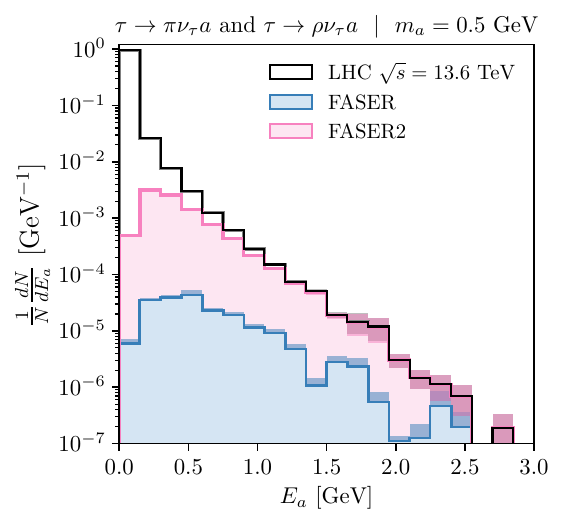}
    \caption{
    The energy distribution of ALPs produced in LFV (top row) and LFC (bottom row) decays of taus at NuMI (left column, $120$~GeV proton beam), SPS (center column, $400$~GeV proton beam), and at the LHC (right column, $\sqrt{s}=13.6$~TeV $pp$ collisions).
    The solid black lines show the total spectrum at the production point and the filled histograms show the respective fraction of the spectrum within the acceptance of each experiment.
    For all panels, we fix $m_a = 0.5$~GeV.
    The uncertainty from the Monte-Carlo statistics is shown as shaded envelopes.
    \label{fig:LFV_spectra}
    }
\end{figure*}

\subsection{ALP decay}
The dominant decay channels of the ALP depend heavily on its mass and the assumed flavor structures of the couplings. 
\cref{fig:alpBRs} shows the dominant branching ratios of a leptophilic ALP as a function of $m_a$ for the LFV anarchy (top), LFC flavor universal (middle), and LFC $e-\tau$ coupled (bottom) benchmark scenarios.
We give the predictions for the individual partial decay widths below.

For an ALP decaying to a pair of leptons of the same flavor the decay width is given by
\begin{align}
	\Gamma(a \to \ell^+\ell^-) = \abs{g_{\ell \ell}}^2\frac{m_\ell^2 m_a}{8\pi f_a^2}\left(1-\frac{4m_\ell^2}{m_a^2}\right)^{1/2},
\end{align}
while for decays to a pair of leptons of differing flavor the partial decay width is given by,
\beq
\begin{split}
	&\Gamma(a \to \ell_1^\pm \ell_2^\mp)
	= \abs{g_{\ell_1\ell_2}}^2\frac{(m_{\ell_1} + m_{\ell_2})^2 m_a}{16\pi f_a^2}	
	 \\
	&~~~~~~~~\times
	\left[1-\frac{(m_{\ell_1}-m_{\ell_2})^2}{m_a^2}\right]
	v(m_a,m_{\ell_1},m_{\ell_2}).
\end{split}
\eeq
Here, we define $a \to \ell_1 \ell_2$ as a sum of both modes, $a \to \ell_1^+ \ell_2^-$ and $a \to \ell_1^- \ell_2^+$, so that 
$\Gamma(a \to \ell_1^\pm \ell_2^\mp)=\Gamma(a \to \ell_1^+ \ell_2^-)+\Gamma(a \to \ell_1^- \ell_2^+)$.
The ALP can also decay into two charged leptons and two neutrinos, \emph{e.g.}, 
$a \to \mu^+ \nu_\tau e^- \bar{\nu}_e$ through $g_{\mu\tau}$.
However, these channels are suppressed by $G_F^{2}$ and by four-body decay phase space factors, and are irrelevant in our discussion.

The ALP--lepton couplings generate at one-loop the $a\to \gamma\gamma$ decays. 
For our analysis these are phenomenologically important only in the corners of parameter space.
Summing all lepton contributions for the $a\to \gamma\gamma$ decay width, we get
\begin{align}
	\Gamma(a\to \gamma\gamma) = \frac{\alpha^2m_a^3}{64\pi^3 f_a^2}
	\abs{\sum_{\ell = e,\mu,\tau} g_{\ell \ell}\left[1-\frac{4m_\ell^2}{m_a^2}I_{\gamma\gamma}^2\left(\frac{m_a}{2m_\ell}\right)\right]}^2,
	\label{eq:a2photon}
\end{align}
where 
\beq
	I_{\gamma\gamma}\left(z\right) = \begin{cases}
	\displaystyle \arcsin\left(z\right) & \mathrm{for}~z \leq 1, \vspace{2mm}\\
	\displaystyle \frac{\pi}{2} + \frac{i}{2}\log\left[\frac{z+\sqrt{z^2-1}}{z-\sqrt{z^2-1}}\right]& \mathrm{for}~z > 1.
	\end{cases}
\eeq
For $m_a \gg 2 m_\ell$, the term with $I_{\gamma\gamma}$ in~\cref{eq:a2photon} becomes much smaller than one. 
For instance, if the axion is $\mu$-phobic then in the range $2m_e<m_a<2m_\tau$ the ratio of the diphoton width to the dielectron width grows with axion mass,
$\Gamma(a\to \gamma\gamma)/\Gamma(a\to ee) \sim \alpha^2 m_a^2/8\pi^2 m_e^2  \sim 8\times (m_a/m_\tau)^2$.  Where we have taken advantage of the numerical coincidence that  $m_e/m_\tau \approx \alpha/8\pi$.  Thus, there is a window around the $\tau$ mass, $m_\tau/3 \lesssim m_a \lesssim 2m_\tau$, where the photon decay mode is dominant.  This can be seen in the bottom panel in \cref{fig:alpBRs}.
For $m_a\ll 2m_\ell$ the photon partial width scales with a high power of the axion mass
\begin{equation}\label{eq:photonwidthbelowlepton}
\Gamma(a\to \gamma\gamma) = \frac{\alpha^2 g_{\ell\ell}^2}{96^2\pi^3}\frac{m_a^7}{m_\ell^4 f_a^2} + \mathcal{O}(m_a^9)~.
\end{equation}
So for light $\tau$-philic axions, where this is the only available decay channel, the axion will also be long lived.

\section{Experimental Signatures}\label{sec:signatures}

\renewcommand{\arraystretch}{1.3}

\begin{table*}[t]
    \centering
    \begin{tabular}{c c c c c c c c}
    \hline \hline
    Experiment & $p^+$ Energy & Exposure &  Baseline & Fiducial Volume & Material(s) & Cuts & Refs. \\
    \hline
\multirow{3}{*}{SHiP} &  & \multirow{3}{*}{$6 \cdot 10^{20}$\,POT}  & \multirow{3}{*}{33.7 m} & 
Pyramidal frustum &  & & \multirow{3}{*}{\cite{Albanese:2878604}} 
\\
& $400$ GeV  & & & ($1\cdot2.7 \to 4\cdot6)\cdot$ & He, & $E_{\ell,\gamma} > 0.5$\,GeV & 
\\
& (SPS) & & & $ 49.6$ m$^3$ & Ecal &  & 
   \\ 
   \hline
   \\
DUNE ND & $120$\,GeV & $11\cdot 10^{21}$\,POT  & $574$~m   & ($5\cdot 5\cdot 5.88$)~m$^3$  & LAr &  $E_{\ell,\gamma} > 0.5$\,GeV  & 
\\
 & (NuMI/LBNF) & (10 yrs) & & & & & \cite{DUNE:2021tad}
\\
\hline
    &  &   & 
677 [723]\,m & $(6\cdot 7\cdot 6)$\,m$^3$ & \multirow{3}{*}{LAr} &  & \multirow{3}{*}{\cite{Coloma:2023adi}} 
\\
protoDUNE & $400$\,GeV & $1.75 \cdot 10^{19}$\,POT  & T2   & $\delta x = + 3 [-3]$\,m,  & &  $E_{\ell,\gamma} > 0.5$\,GeV  & 
\\
NP02 [NP04] & (SPS) & (5 yrs) &  & $\delta y = +1.5$\,m & & & 
\\
\hline
\multirow{3}{*}{FASER} &  & \multirow{3}{*}{$\mathcal{L} = 150$ fb$^{-1}$} & \multirow{3}{*}{480\,m} & $R=0.1$\,m, & &  & \multirow{3}{*}{\cite{FASER:2023tle}} 
\\
& $\sqrt{s} = 13.6$\,TeV & &  
 & $z=1.5$\,m &  Air, &  $E_{\ell,\gamma} >50$\,GeV &
\\
& (LHC) &  
&  & $\delta y =6.5$\,cm & Ecal &  &
\\
    \hline
    \multirow{2}{*}{FASER2} & $\sqrt{s} = 13.6$\,TeV & \multirow{2}{*}{$\mathcal{L} = 3$ ab$^{-1}$} & 

  \multirow{2}{*}{650 m} & \multirow{2}{*}{$2.6\cdot 1\cdot 10$ m$^3$} & Air, &   \multirow{2}{*}{$E_{\ell,\gamma} > 50$\,GeV} & \multirow{2}{*}{\cite{FPFWorkingGroups:2025rsc}} 
\\
&(LHC) & & 
& & Ecal &  &
    \\
    \hline
    \multirow{2}{*}{ICARUS} & $120$\,GeV  & \multirow{2}{*}{$3\cdot 10^{21}$\,POT}  & 803 m & 
$2\times(2.67\cdot $ & \multirow{2}{*}{LAr} & \multirow{2}{*}{$E_{\ell,\gamma} > 0.1$ GeV} & \multirow{2}{*}{\cite{Batell:2019nwo,ICARUS:2024oqb}} 
\\
& (NuMI) & &($5.6^\circ$) & $\cdot 2.86) \cdot 17.0$m$^3$ & & &
    \\
\hline
    \multirow{2}{*}{NOvA ND} & $120$\,GeV  & $4.2\cdot 10^{21}$\,POT  & 990 m & $3.9\cdot 3.9 \cdot$ & Liquid & \multirow{2}{*}{$E_{\ell,\gamma} > 0.1$ GeV} & \multirow{2}{*}{\cite{NOvA:2007rmc}} 
\\
& (NuMI) & (2024)  & (14.6 mrad) & $\cdot 12.7$ m$^3$ & Scint. & &
   \\
\hline
    \multirow{2}{*}{MicroBooNE} & $120$\,GeV  &  \multirow{2}{*}{$2\cdot 10^{21}$\,POT}  & 685 m & $2.26\cdot 2.03 \cdot$ &  \multirow{2}{*}{LAr} & \multirow{2}{*}{$E_{\ell,\gamma} > 0.1$ GeV} & \multirow{2}{*}{\cite{MicroBooNE:2021sov}} 
\\
& (NuMI) &  & ($8^\circ$) & $\cdot 9.42$ m$^3$ &  & &
   \\
\hline
\multirow{2}{*}{BEBC} & $400$\,GeV & \multirow{2}{*}{$2.72\cdot 10^{18}$\,POT}  & \multirow{2}{*}{404 m} & $3.57\cdot 2.52\cdot$ & \multirow{2}{*}{Ne/H$_2$} & \multirow{2}{*}{$|\vec{p}_{\ell,\gamma}| > 0.5$ GeV}  & \multirow{2}{*}{\cite{WA66:1985mfx,Barouki:2022bkt}} 
\\
& (SPS) & &  & $\cdot 1.85$ m$^3$ & &  &
\\
\hline
    \multirow{2}{*}{CHARM} &$400$ GeV & \multirow{2}{*}{$2.4\cdot10^{18}$\,POT}  & \multirow{2}{*}{480 m} & $ 3 \cdot 3\cdot 35$ m$^3$ & Air, &  \multirow{2}{*}{$E_{\ell,\gamma} > 0.5$ GeV} & \multirow{2}{*}{\cite{CHARM:1985anb,CHARM:1985nku}} \\
& (SPS) & & & $\delta x = 5$~m & Ecal & & 
\\
    \hline \hline
    \end{tabular}
    \caption{A list of experiments considered in this work and their respective parameters.
    Here, $\mathcal{L}$ stands for the assumed LHC luminosity, and $\delta x (\delta y)$ are the assumed locations of the centers of the detectors in the horizontal (vertical) directions.  The dimensions of the detector are given as width$\cdot$height$\cdot$depth and the baseline is the distance from the point of $\tau$ production to the front of the detector.  Experiments for which the axion decay products are required to hit an electromagnetic calorimeter at the back of the fiducial volume are denoted with ``Ecal".}
    \label{tab:experiments}
\end{table*}


Next, we describe how we calculate the expected tau production rate and the resulting ALP signals at different experiments.

\subsection{Tau production rate}
\label{sec:tau_production}

We are particularly interested in the possible ALP fluxes from tau decays at three different proton beams:
\begin{enumerate}
    \item NuMI at Fermilab: $120$~GeV protons on a fixed target,
    \item SPS at CERN: $400$~GeV protons on a fixed target,
    \item LHC: $\sqrt{s} = 13.6$~TeV center of mass (COM) energy of $pp$ collisions.
\end{enumerate}
Using \textsc{Pythia\,8}~\cite{Bierlich:2022pfr} (\textsc{v8.310} with the Monash tune choice of parameters~\cite{Skands:2014pea}), we simulate individual $pp$ collisions at each of the above three beam--target configurations.
We checked that simulating $pn$ collisions at fixed-target experiments gives similar $\tau$yields and distributions.

We are interested primarily in the forward production of $D_{(s)}^\pm$, $\psi(2S)$, as well as direct $\tau^\pm$ production (the latter, though, is numerically subleading).  
Thus, we use the \texttt{SoftQCD:all} flag to simulate their production and turn off all \texttt{HardQCD} processes to avoid double counting.
In each setup, we track the $D^\pm$ and $D_s^\pm$ mesons and $\psi(2S)$ charmonium produced by \textsc{Pythia} and force them to decay to $\tau^\pm$ leptons, weighting each event by the appropriate branching ratio to taus.

To speed up our simulations, we fit the transverse momentum $p_T$ and Feynman $x_F$ distributions of taus to the \pythia output using empirical analytical formulae.
With the analytical fits, we can generate weighted tau events more efficiently, covering kinematical tails of distributions with smaller number of events and speeding up our parameter scan significantly.
We do this for $\tau^+$ and $\tau^-$ separately for all three beam setups above.
Our fits also provide a simple description of tau production at the three beams that can be used in future applications.

In each detector, we load the same $\tau^\pm$ events generated with the analytical formulae and force taus to decay to ALPs for each of the new-physics decay channels discussed in \cref{sec:ALP:prod:LFV,sec:ALP:prod:LFC}, keeping track of the weights and the corresponding branching ratios to ALPs.

In the estimates we neglect the interaction of secondaries within the target.
While this component has been argued to produce a non-negligible fraction of charmed mesons and tau particles in thick targets~\cite{Coloma:2020lgy}, it does not invalidate the bounds derived here.
On the contrary, including this component can only boost the number of taus produced --- our regions of coverage can be viewed as being conservative.

In our simulations we obtain the following inclusive $\tau^\pm$ yields per proton--nucleon ($pN_T$) collision,
\begin{enumerate}
    \item NuMI: $2.9\times10^{-7}$ $\tau^\pm$/$pN_T$;
    \item SPS:  $1.7\times10^{-6}$ $\tau^\pm$/$pN_T$;
    \item LHC: tau production cross section $\sigma^\text{tot}_\tau= 0.22~\text{mb}$.
\end{enumerate}
\Cref{fig:tau_spectra} displays the kinematic distributions of the produced $\tau^\pm$ leptons. 
The simulated samples correspond to approximately $2.2\times10^{6}$ $\tau$'s for NuMI, $6.1\times10^{6}$ for SPS, and $2.1\times10^{6}$ for the LHC.
In \cref{app:tauprod}, we discuss potential uncertainties in these predictions and their impact on the experimental sensitivities presented here.

We note that the issue of estimating the number of ALPs from tau decays at beam dump experiments is similar in nature to the calculation of the $\nu_\tau/\overline\nu_\tau$ flux.
Both depend primarily on the production of $D_s^\pm$ mesons, and therefore, on an accurate description of open-charm production. 
At SHiP, there have been several estimates of this flux over the years~\cite{Alekhin:2015byh,Bai:2018xum,Maciula:2019clb,FASER:2024ykc}, with total yields that vary by as much as an order of magnitude.
In the future, the yields can be experimentally determined with significantly improved precision using data from the DsTau experiment (NA65 at the CERN SPS)~\cite{DsTau:2019wjb}, SHiP-charm~\cite{SHiP:2024oua}, and with the neutrino data from SHiP itself~\cite{Alekhin:2015byh}, reducing the uncertainty on the ALP flux predictions.
At FASER, the situation is similar and $\nu_\tau/\overline\nu_\tau$ data will help constrain it, as it already does for $\pi,K$ production using $\nu_\mu/\overline\nu_\mu$ data~\cite{FASER:2024ref}.
The lower COM energy at the NuMI beam means that there the uncertainty on the ALP flux is expected to remain large, given that the flux of $\nu_\tau/\overline\nu_\tau$ is too small to be observed.
For more detailed studies on the impact of charm and tau production on new physics, see Refs. \cite{Fieg:2023kld} and \cite{Schubert:2024hpm}.

\subsection{ALP flux and decays in flight}
\label{sec:alpflux}

The ALP energy spectrum $E_a$ in the tau rest frame is obtained by sampling the differential decay rates discussed in \cref{sec:ALPprod}.
The ALP we consider here couples exclusively via the axial-vector current, so that its emission in the tau rest frame is isotropic, independent of the tau polarization~\cite{Calibbi:2020jvd}.
In models where the ALP couples to left- or right-handed currents, then the decay is no longer isotropic for polarized taus.
However, even in that case, the effect of the angular-dependence is small as the net polarization of the $\tau^+ + \tau^-$ produced at the target is small due to $\tau^+/\tau^-$ production asymmetry being  small. 
We find that the ratio of $\tau^+$ to $\tau^-$ production rates at NuMI, SPS, and LHC is $0.73$, $0.84$, and $0.99$, respectively.
The dominance of $D_s^-$ over $D_s^+$ at low energies is consistent with leading particle effects in charm hadroproduction~\cite{E769:1996xad}.

The next step in our simulation is to calculate the expected number of decays-in-flight at different detectors.
We assume that the detector's face is aligned with the line-of-sight to the target and then calculate the probability $P_{\rm decay} $ of an ALP to decay to a given final state $a \to \ell_1 \ell_2$ inside the fiducial volume as
\beq
\label{eq:Pdecay}
    P_{\rm decay} = \mathcal{B}(a \to \ell_1 \ell_2) \exp\left[- \frac{\Gamma_a L}{\gamma_a \beta_a}\right]
    \left[1 -\exp\left[-\frac{ \Gamma_a\lambda}{\gamma_a \beta_a} \right]\right],
\eeq
where $\Gamma_a$ is the total ALP decay width in the rest frame, $\gamma_a$ and $\beta_a$ are the boost factor and speed of the ALP, $L$ is the distance from the target to the face of the detector, and $\lambda$ is the distance that each ALP travels inside the fiducial volume.
For long-lived ALPs ($\Gamma_a L \ll 1$ and $\Gamma_a \lambda \ll 1$), the expression in \cref{eq:Pdecay} simplifies to $P_{\rm decay} =\lambda \Gamma_{a \to \ell_1 \ell_2}/\gamma_a \beta_a$, where $\Gamma_{a \to \ell_1 \ell_2}$ is the partial decay width for the $a\to \ell_1\ell_2$ channel.  
For experiments whose decay volume is not instrumented (CHARM, SHiP, FASER and FASER-2, see \cref{tab:experiments}), we furthermore estimate the probability for the decay products to reach the far side of the decay volume, after which these are assumed to be detected in the calorimeter. 
In all cases we apply simple energetic requirements on the final state leptons/photons.  Beyond these geometric requirements -- that axions enter the front of the fiducial volume, decay in the detector volume, and the leptons hit the electromagnetic calorimeter (for the experiments that require it) -- and simple energetic cuts, we do not apply detector efficiencies.

\subsection{ALP detection}

The list of detectors we consider at the three locations is shown in \cref{tab:experiments}.
For each, we quote the assumed fiducial volume, the position of the center of the detector face, and their known or expected exposure.
For FASER and FASER-2, we use the results from~\cite{Barr_FASER2_2025}.
While a limit on ALP couplings from FASER data could in principle already be placed by recasting their latest search for dark photons~\cite{FASER:2023tle}, we find that this is not yet competitive with other direct limits.
For ProtoDUNE, we follow closely Ref.~\cite{Coloma:2023adi}.
The faces of the two detectors, NP02 and NP04, are not aligned with the direction of the beam. We expect that accounting for this misalignment would not significantly affect our results.
We take into account $\tau$ production in the primary target and neglect the one from the beam dump, where a smaller, less-energetic fraction of the beam stops.
For ICARUS and MicroBooNE, we follow Refs.~\cite{Batell:2019nwo,ICARUS:2024oqb}, assuming a total NuMI POT of $2\times 10^{21}$ at MicroBooNE~\cite{MicroBooNE:2025prw} and a projected $3\times 10^{21}$ POTs at ICARUS.
We note that the polarity of the magnetic horns is not important for the ALP signal since it is produced by the promptly-decaying $D_{(s)}$ and $\tau$ particles, and so we sum the so-called forward and reverse horn current data taking periods.

For the DUNE near detector (ND), one may consider a few different possibilities: ND LAr, ND GAr, or SAND, each with their unique set of advantages.
For charged final states, ND GAr will likely be preferable from the point of view of backgrounds and efficiencies.
For two-photon decays, however, a higher density detector would be needed, so NDLAr or SAND would be more relevant.
In \cite{Coloma:2023oxx}, the authors find large detection efficiencies for ND GAr, above $90\%$ for the GAr module.
For SHiP, we take the numbers from~\cite{Albanese:2878604}, quoting the sensitivity for the benchmark of $6\times 10^{20}$~POTs.

Our results are quoted in terms of event-rate sensitivities, \emph{i.e.}, the regions where experiments can expect more than 2.3 signal events during their total exposure.
In a background-free search, this would correspond to the $90\%$~CL sensitivity.
The zero (or $\ll 1$) background limit is applicable to the existing searches at CHARM and BEBC. In these two cases one can therefore interpret our results as the $90\%$~CL exclusions.
For future searches, whether or not the zero background limit will apply will depend on the details of the experiments and the analysis strategies adopted. 
For instance, in the existing searches for dilepton pairs at MicroBooNE~\cite{MicroBooNE:2021sov,MicroBooNE:2022ctm,MicroBooNE:2023eef} and ICARUS~\cite{ICARUS:2024oqb}, requiring the dilepton pair invariant mass to coincide with the assumed axion mass, reduces backgrounds to be below or at the level of one event, with efficiencies that range from $10\%-50\%$.
The ALP signals discussed here are of more energetic than, for instance, the signals one would obtain from LLPs that would be produced in decays of kaons at rest. 
As such, we expect the neutrino-induced backgrounds to be less important in the case of ALPs from tau decays.

On the other hand, note however, that the reconstruction of dimuon pairs may worsen compared to LLPs, since the muons are now more likely to escape the fiducial volume, although it opens up the possibility to search for non-starting ALP decay events.
Understanding how big this effect is, however, requires a rather detailed analysis of the backgrounds and the selections, which we expect to be done by the collaborations as part of their search strategy optimization.

\subsection{Other experiments}

Let us also comment on several potentially relevant experiments that we do not include in the figures.
The ArgoNeuT detector at the Fermilab MINOS near detector hall has searched for heavy neutral leptons and axions-like-particles using dimuon pairs starting outside or inside of its Liquid Argon Time Projection Chamber (LArTPC) fiducial volume~\cite{ArgoNeuT:2021clc,ArgoNeuT:2022mrm}.
In Ref.~\cite{Bertuzzo:2022fcm}, the authors adapted these searches to place new limits on the parameter space of leptophilic ALPs and Majorons.
We take $87\%$ of ALPs to be produced at the NuMI target and $13\%$ at the location of the beam dump, counting as signal all ALP decays to dimuons that take place within a $40\times 47 \times 210$~cm$^3$ volume that starts $1033$~m away from the target and is centered around the proton beam.
With a simplified analysis which takes the effect of geometric cuts and signal efficiency to be independent of energy and tuning their overall effect $\epsilon$, we find some agreement with Bertuzzo et al when $\epsilon\approx 0.5$ but were unable to fully reproduce their results.  We find some agreement with the limits on $1/f_a$ from Ref.~\cite{Bertuzzo:2022fcm} for the case $R_a = g_{\ell \ell} / g_{\ell_1 \ell_2} = 5$, but not for $R_a = 1/3$.
Nevertheless, we find that ArgoNeuT is not competitive with existing limits placed by CHARM and BEBC searches, even when using the exclusion curves in Ref.~\cite{Bertuzzo:2022fcm}, or when using the most optimistic assumptions about the efficiencies and exposure.

Recently, the $2\times2$ DUNE near detector demonstrator, also a LArTPC detector, has been operating in the MINOS hall~\cite{Russell:2024svv,Wresilo:2024zol}.
It consists of four tall detectors, each $0.7\times1.4\times 0.7$\,m$^3$ of volume and located at a similar distance to the NuMI target as ArgoNeuT.
We find that for $100\%$ efficiencies and no backgrounds, a search for starting events (decays inside the LArTPCs) would not be competitive with existing limits or with the event-rate sensitivities of MicroBooNE, ICARUS, and NOvA even for an exposure of $10^{21}$~POTs.

The NA62 kaon detector has also been used to set limits on long-lived particles using the SPS beam~\cite{NA62:2023nhs,NA62:2025yzs}.
We estimate that in the future, with $10^{18}$ POT~\cite{Jerhot:2936260}, NA62 in beam dump mode may outperform existing CHARM and BEBC limits on the ALP models we are interested in.
Finally, the past NuCal detector at the $70$~GeV proton beam at IHEP could also be potentially relevant~\cite{Blumlein:1990ay,Blumlein:1991xh}.
While we do not study its reach in detail, we observe that the smaller $\tau^\pm$ yield due to the lower energy of the beam could be compensated by the closer location of the detector ($63$~m) and its relatively large fiducial volume ($2.4\times2.4\times23$\,m$^3$) and $1.7 \times 10^{18}$~POTs (see, for instance,~\cite{Tsai:2019buq,Kyselov:2024dmi}).

\section{Results} 
\label{sec:elimits}

\begin{figure*}[t!]
    \centering
    \includegraphics[width=0.49\textwidth]{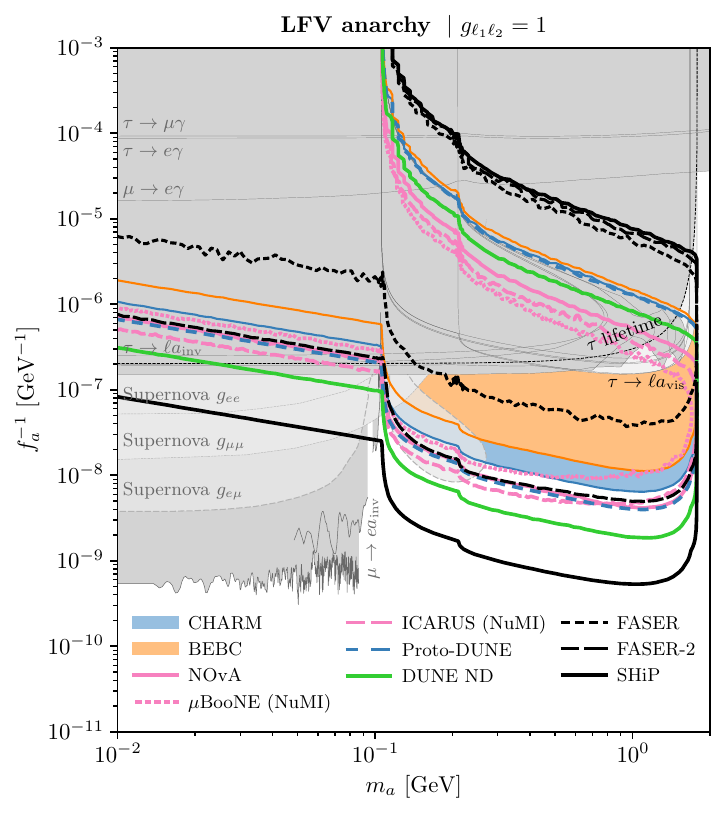}
    \includegraphics[width=0.49\textwidth]{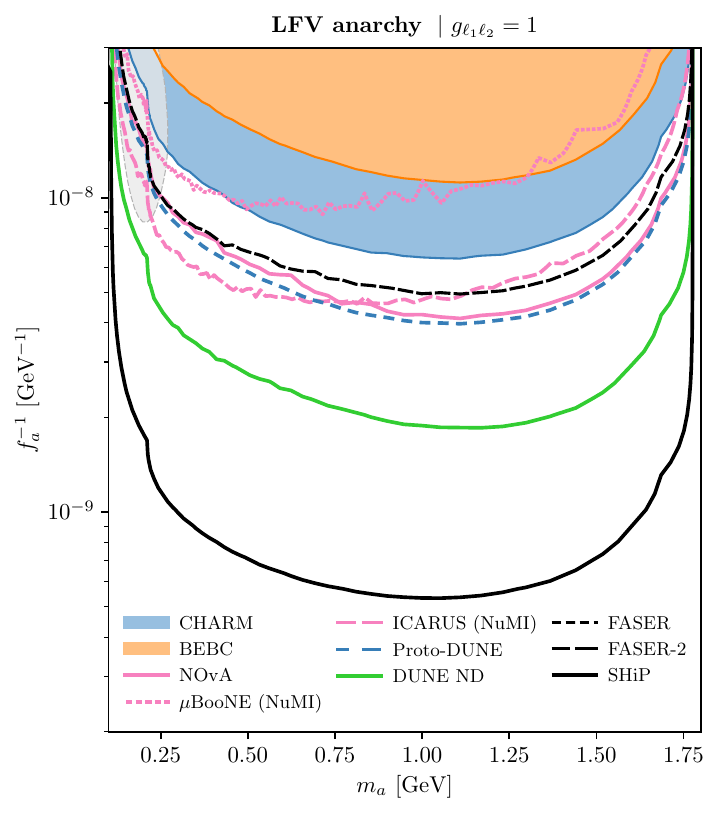}
    \caption{The parameter space of flavor-violating ALPs with flavor anarchical couplings, \cref{eq:flavor_anarchy}. 
    The right panel shows a zoomed-in version of the parameter space shown in the left panel.
    Constraints on long-lived ALPs discussed in \cref{sec:signatures} are shown as filled regions for existing limits from CHARM and BEBC and as solid and dashed lines for projections for neutrino experiments, FASER, proto-DUNE, and SHiP.
    Other constraints from tau decays, discussed in \cref{sec:elimits}, are shown as light grey regions for $\tau \to \ell a_{\rm inv}$ and dark grey regions for $\tau \to \ell a_{\rm vis}$.
    \label{fig:LFV_anarchy}
    }
\end{figure*}

\begin{figure*}[t!]
    \centering
    \includegraphics[width=0.49\textwidth]{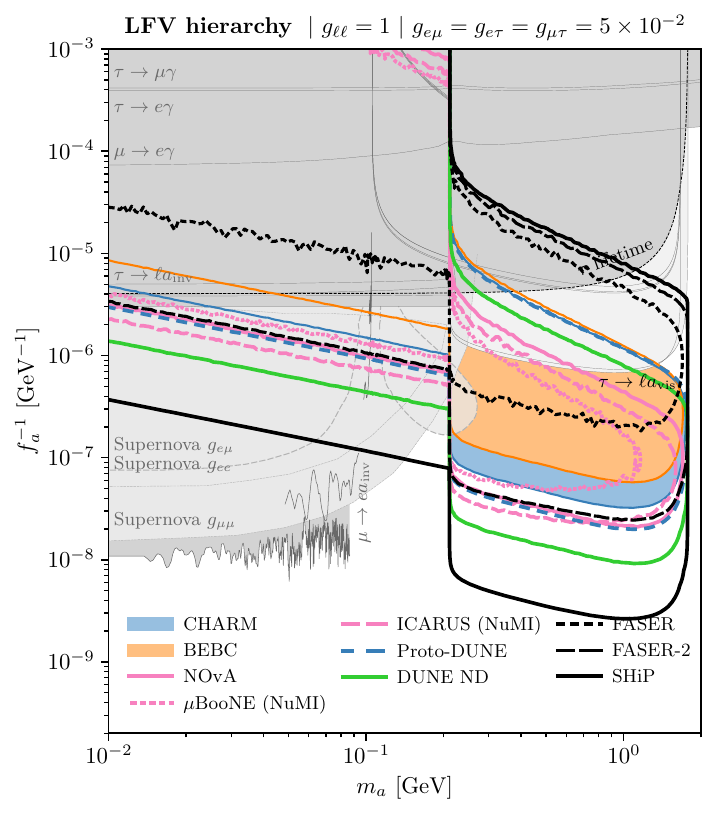}
    \includegraphics[width=0.49\textwidth]{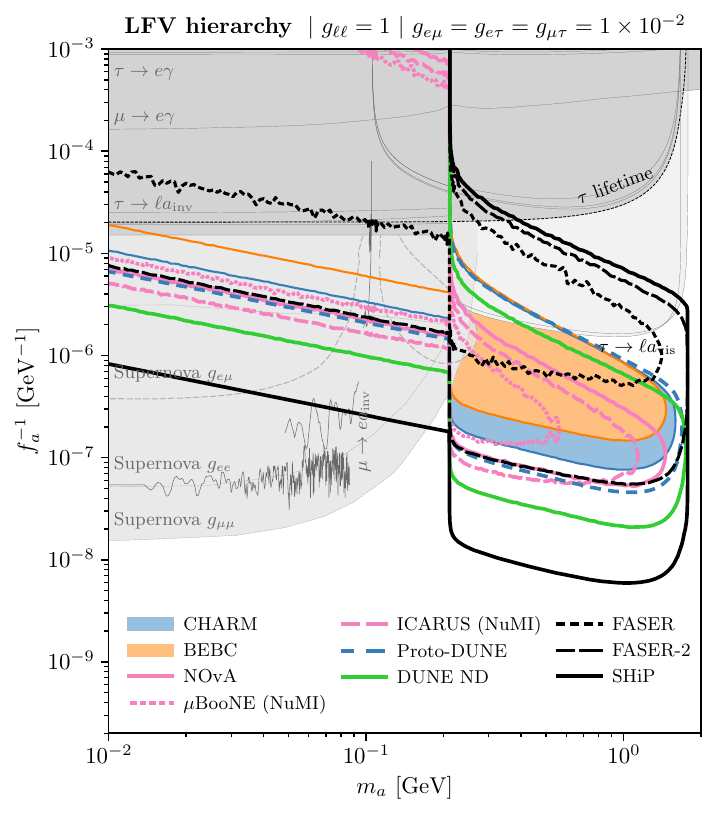}
    \includegraphics[width=0.49\textwidth]{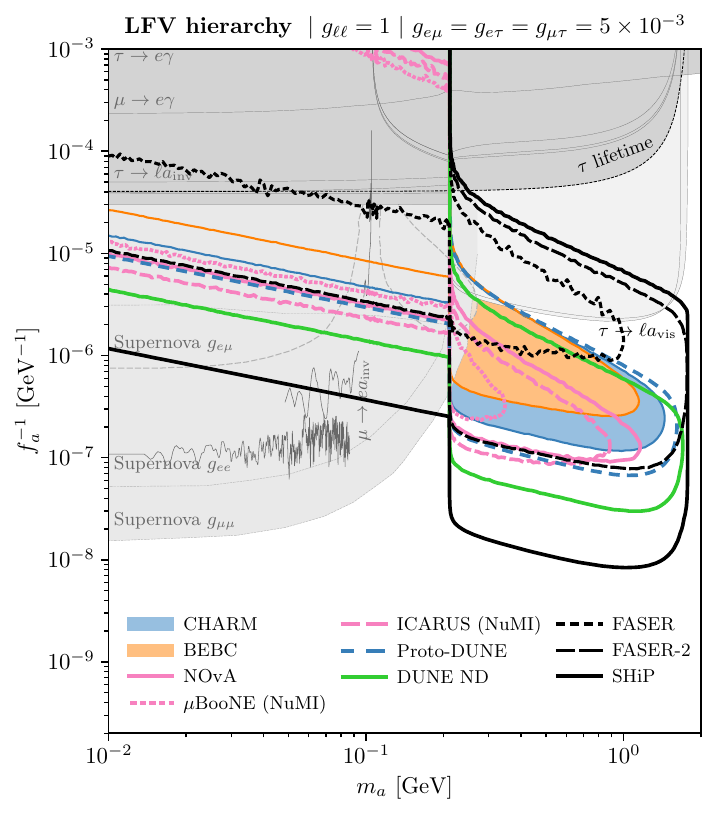}
    \includegraphics[width=0.49\textwidth]{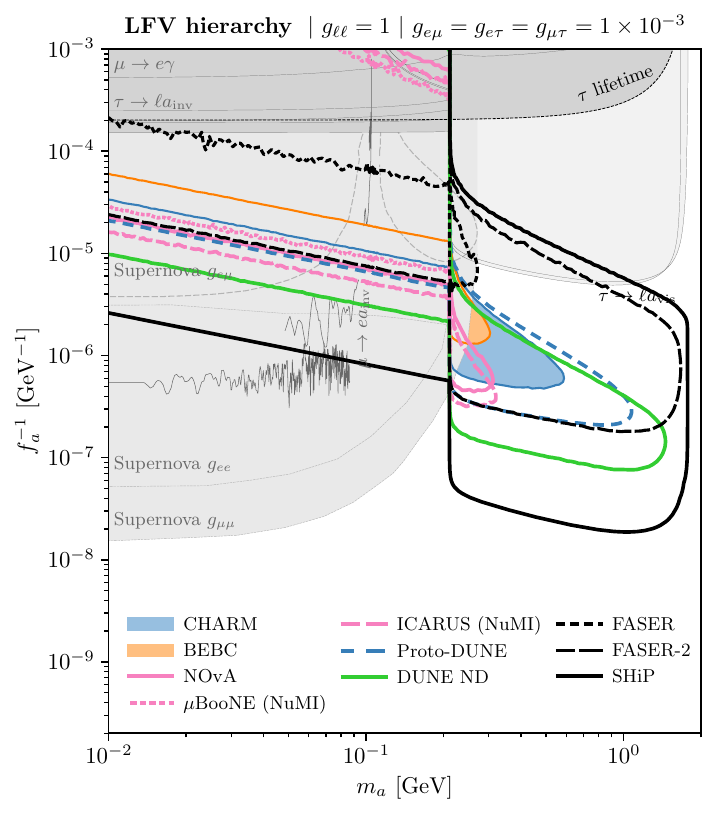}
    \caption{Same as \cref{fig:LFV_anarchy}, but for flavor-violating hierarchical couplings, \cref{eq:flavor_hierarchy}, for several representative sizes of flavor off-diagonal couplings, as indicated.
    Constraints on long-lived ALPs discussed in \cref{sec:signatures} are shown as filled colored regions for existing limits from CHARM and BEBC and as solid and dashed lines for projections for neutrino experiments, FASER, proto-DUNE, and SHiP.
    Other constraints from tau decays, discussed in \cref{sec:elimits}, are shown as light grey regions for $\tau \to \ell a_{\rm inv}$ and dark grey regions for $\tau \to \ell a_{\rm vis}$.
    \label{fig:LFV_hierarchy}
    }
\end{figure*}

\begin{figure}[t!]
    \centering
    \includegraphics[width=0.49\textwidth]{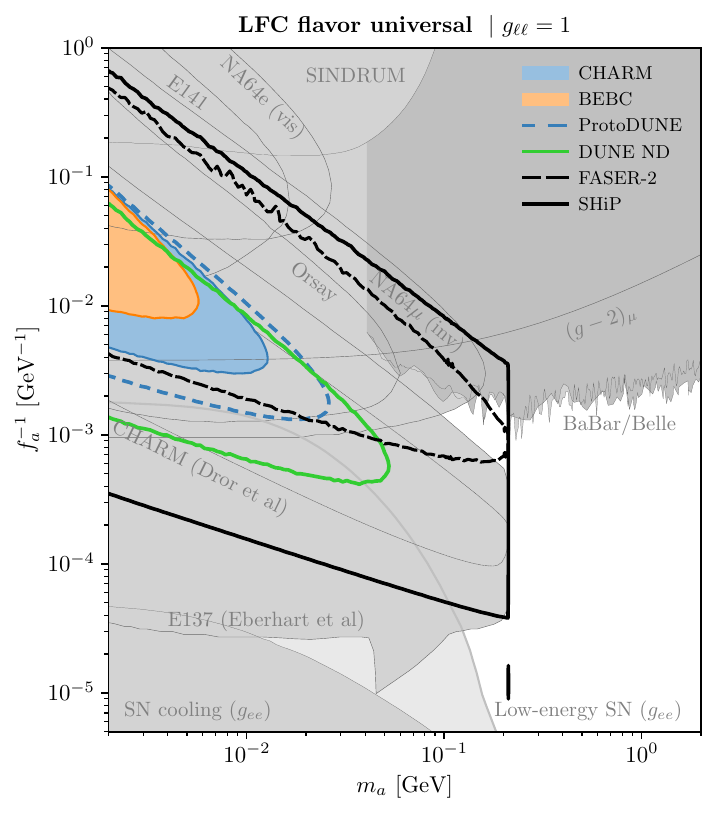}
    \caption{Limits and projected event-rate sensitivities in the parameter space of LFC ALP with universal couplings of \cref{eq:flavor_conserving_universal}.
    Constraints on long-lived ALPs discussed in \cref{sec:signatures} are shown as filled regions for existing limits from CHARM and BEBC and as solid and dashed lines for projections for FASER-2, proto-DUNE, DUNE, and SHiP.
    Other neutrino experiments and FASER are less sensitive and not shown.
    The grey regions show limits from SINDRUM~\cite{SINDRUM:1986klz} and CHARM~\cite{CHARM:1985anb} as derived in \cite{Altmannshofer:2022ckw}, electron beam dumps limits derived in as derived in~\cite{Liu:2016qwd} and \cite{Araki:2021vhy} (see also~\cite{Bjorken:2009mm,Andreas:2010ms}) with data from E137~\cite{Bjorken:1988as}, E141~\cite{Riordan:1987aw}, and Orsay~\cite{Davier:1989wz}, as well as limits from BaBar~\cite{BaBar:2020jma}, Belle~\cite{Belle:2022gbl}, and NA64~\cite{NA64:2021aiq}.
    We also show supernova limits based on cooling arguments~\cite{Carenza:2021pcm,Fiorillo:2025sln} as well as the constraint from ALP energy depositions in the outer envelope of the supernova~\cite{Caputo:2022mah,Fiorillo:2025sln} as a lighter grey line.
    \label{fig:LFC}
    }
\end{figure}

\begin{figure*}[t!]
    \centering
       \includegraphics[width=0.49\textwidth]{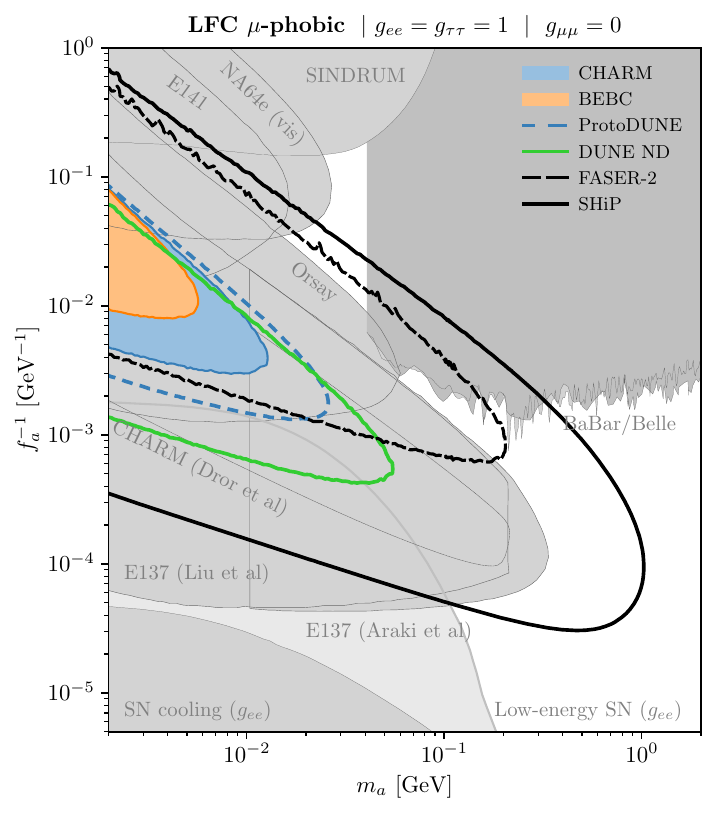}
    \includegraphics[width=0.49\textwidth]{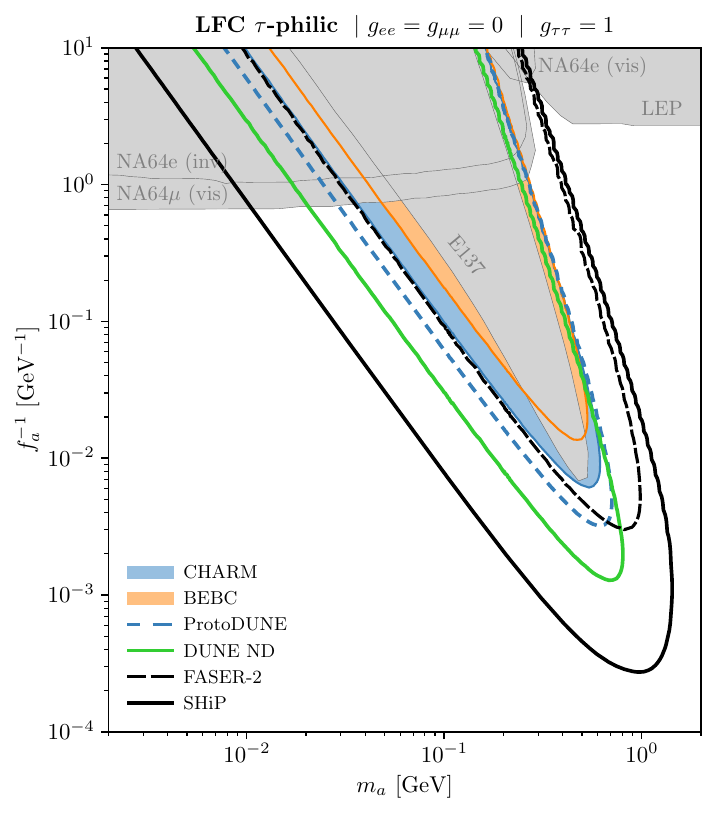}
    \caption{Same as \cref{fig:LFC} but for two flavor nonuniversal benchmarks a $\mu$-phobic ALP (left) and a $\tau$-philic ALP (right), cf.~\cref{eq:flavor_conserving_nonuniversal,eq:tau:philic}.
    The grey regions in the right panel show the limits derived in \cite{Eberhart:2025lyu} using data from E137~\cite{Bjorken:1988as}, NA64~\cite{Gninenko:2719646,NA64:2019auh,NA64:2024klw}, and LEP~\cite{Jaeckel:2015jla}.
    \label{fig:LFC_tauphilic}
    }
\end{figure*}

Next, we present our results for the experimental reach for each of the benchmark LFV and LFC ALP scenarios introduced in~\cref{sec:LLP}, see~\cref{eq:flavor_anarchy,eq:tau:philic}.\footnote{
In the LFV anarchy and LFC universal cases, we have checked that our results are in reasonable agreement with \textsc{SensCalc}~\cite{Ovchynnikov:2023cry,Kyselov:2024dmi}
when using the same $\tau$ yields obtained in our simulation.
We note that \textsc{SensCalc} includes the secondary charm production at SPS and therefore uses different $\tau$ distributions,
and approximates the four-body $\tau$ decay matrix element to be constant.
}

\subsection{Flavor-violating ALPs}

For LFV ALPs, we consider two benchmarks, the anarchic flavor pattern,~\cref{eq:flavor_anarchy}, and the flavor hierarchical case,~\cref{eq:flavor_hierarchy}.
The exclusions and the projected event-rate sensitivities for the flavor anarchy LFV case, \cref{eq:flavor_anarchy},  are shown in \cref{fig:LFV_anarchy}.
The left panel shows as filled colored regions in the $(m_a, f_a^{-1})$ plane the bounds we derived from CHARM and BEBC searches.
These are compared to other direct limits on ALPs from muon and tau decays as well as supernovae, shown in shades of grey (see \cref{sec:otherlimts}).
The projected event-rate sensitivities of various experiments are shown as lines of different colors and styles, as indicated in the legend.

The right panel in \cref{fig:LFV_anarchy} shows a zoomed-in region of the parameter space.
Despite the lower energy of the Main Injector beam compared to SPS and the LHC, neutrino detectors off-axis from the NuMI target, namely NOvA, MicroBooNE and ICARUS, may improve on existing limits, if a high-efficiency, low-background search can be performed.
The SBND detector could, in principle, also be considered, however, it is much further off-axis than the other three.
We note that in the models we are interested in the ALP signatures are typically of relatively high-energy (see \cref{fig:LFV_spectra}), so that a search for a non-starting dimuon signal could be advantageous, further increasing the effective fiducial volume of the experiment, as already demonstrated by the ArgoNeuT experiment in Refs~\cite{ArgoNeuT:2021clc,ArgoNeuT:2022mrm}.

\Cref{fig:LFV_hierarchy} shows our results for the flavor hierarchical cases,~\cref{eq:flavor_hierarchy}.
Each panel corresponds to a different hierarchy between LFV and LFC couplings;
the flavor diagonal couplings are set to $g_{\ell \ell} = 1$, while the values of off-diagonal couplings $g_{\mu \tau}$, $g_{e\tau}$, and $g_{e\mu}$ decrease from top left to bottom right panel. For simplicity, all off-diagonal couplings were set to be equal, $g_{\mu \tau}=g_{e\tau}=g_{e\mu}$, in the four representative cases shown. That is, the variation of LFV couplings is modeled using a single parameter $\lambda \ll 1$, setting,
\begin{equation}
\label{eq:LFV:ansatz}
    \left(\begin{matrix}
        g_{ee} & g_{e \mu} & g_{e\tau} \\
        g_{e\mu} & g_{\mu \mu} & g_{\mu\tau} \\
        g_{e \tau} & g_{\mu \tau} & g_{\tau \tau} \\
    \end{matrix}\right) = 
     \left(\begin{matrix}
        1 & \lambda & \lambda \\
        \lambda & 1 & \lambda \\
        \lambda & \lambda & 1   \\
    \end{matrix}\right).
\end{equation}
The decreasing values of off-diagonal couplings in the panels in \cref{fig:LFV_hierarchy} illustrate the effect of  a decreasing amount of LFV on the ALP phenomenology. 
In practice, this variation amounts to changing the ALP production rates without substantially modifying the ALP lifetime, the latter being driven by $g_{ee,\mu\mu}$.
Consequently, the experiments that are sensitive to shorter-lived ALPs, such as FASER and SHiP, become increasingly more relevant.

From low energy perspective, the exact choice for the values of the off-diagonal ALP couplings in \cref{eq:LFV:ansatz} is arbitrary, since it depends on the origin of flavor structure and thus on details of UV physics. Changing the relative sizes of $g_{e\mu}$ and $g_{e(\mu)\tau}$ couplings does not change appreciably the projected sensitivities of beam-dump like experiments. It does, however, directly control the importance of CLFV searches. 
For the choice of equal size LFV couplings, $g_{e\mu} = g_{e(\mu)\tau}$, the constraints from $\mu \to ea$ decays are typically the dominant laboratory constraints whenever $m_a < m_\mu - m_e$.
This need not be the case if, for instance, we had assumed $g_{e\mu} \ll g_{e(\mu)\tau}$.
Nevertheless, for the LFV hierarchical case, the choice of $g_{e\mu}$ has little impact on ALP searches at high-energy beam dumps in the most relevant mass region, namely $m_a > 2 m_\mu$.
In the intermediate region of $m_e + m_\mu < m_a < 2m_\mu$, the LFV ALP decay channel can be important depending on the size of $g_{\mu e}/g_{ee}$.

\subsection{Flavor-conserving ALPs}

The left panel of \cref{fig:LFC} shows our results for the flavor-conserving scenario,~\cref{eq:flavor_conserving_universal}, with universal couplings to all three generations of SM leptons, $g_{ee} = g_{\mu \mu} = g_{\tau \tau}$. For ALP production due to ALP bremsstrahlung modifying the SM $\tau$ decays,
the additional $G_F$ suppression of the decay amplitude compared to the LFV case
 means that the sensitivity to scale $f$ is significantly weaker.
Existing limits from CHARM and BEBC based on ALP production in tau decays are comparable with the other beam dump limits shown in various shades of grey.
The latter are based on electron bremsstrahlung and light meson decays. Given that the larger number of electrons and mesons available at the beam dump facilities, these searches can still be relevant to the LFC scenarios, despite the strong suppression of ALP couplings to electrons.
We find that the other considered experiments, MicroBooNE, ICARUS, NOvA, and FASER, are not competitive with the electron-based limits.
In the future, however, the high-energy experiments such as FASER-2 and SHiP will be able to explore new parameter space using only ALP production from the tau decays we calculate here.

 \Cref{fig:LFC_tauphilic} shows the results for the two flavor non-universal LFC benchmarks, $g_{ee}=g_{\tau\tau}=1, g_{\mu\mu}=0$ ($\mu-$phobic, left panel), and $g_{ee}=g_{\mu\mu}=0, g_{\tau\tau}=1$ ($\tau$-philic, right panel).
In both benchmarks the ALP does not couple to muons, which has important phenomenological consequences.  
For instance, the flavor universal LFC ALP, \cref{fig:LFC}, with a mass above the dimuon threshold, $m_a > 2 m_\mu$, would decay via $a\to \mu^+\mu^-$ with a lifetime too short to be observed in beam dump experiments.
For vanishing coupling to muons, $g_{\mu \mu} = 0$, the ALP is long-lived also for masses above the dimuon threshold, decaying either via  $a\to e^+e^-$ or $a\to \gamma\gamma$, and thus beam dump searches are sensitive to the two scenarios also above the muon threshold. 
In fact, for a $\tau$-philic ALP with $g_{ee} = g_{\mu \mu } = 0$, only $a \to \gamma \gamma$ decays are possible below the di-tau threshold, $m_a < 2m_\tau$, making the $\tau$-philic benchmark significantly less constrained than the $\mu$-phobic one.  Note also,  that in the $\tau$-philic case there is a steeper dependence of the sensitivities to ALP mass because the photon width scales with a high power of ALP mass, $\Gamma(a\to \gamma\gamma)\sim m_a^7$, see~\cref{eq:photonwidthbelowlepton}. 
In summary, as shown in~\cref{fig:LFC_tauphilic}, ALP production from tau decays at beam dumps are again providing some of the leading constraints in the parameter space also for the flavor non-universal LFC ALPs.

\subsection{Existing limits}
\label{sec:otherlimts}

In \cref{fig:LFV_anarchy,fig:LFV_hierarchy,fig:LFC,fig:LFC_tauphilic} we compare the reach of beam dump experiments with other probes of weakly-interacting particles.
Because the ALP couples exclusively to leptons, many limits on long-lived ALPs from pion or kaon decays, such as~\cite{ArgoNeuT:2022mrm,Coloma:2022hlv,Berger:2024xqk}, are too weak to appear in our plots.
This is particularly true for LFV couplings, where the ALP production in fully leptonic 2-body $\tau$ and $\mu$ decays is significantly enhanced relative to any $G_F$-suppressed production mode from pion and/or kaon decays.
The tau lifetime measurements, bounds on some of  the rare tau decay channels, and the astrophysical constraints therefore pose more meaningful limits on the ALP decay constant.
For the LFC case, the tau and muon ALP-branching ratios are suppressed by $G_F$ and are thus much smaller than in the LFV case for the same value of $f_a$.
In this case direct constraints, therefore, come mostly from ALP couplings to electrons.

\emph{Two-body decays of taus} provide some of the leading direct constraints on the lepto-philic ALPs.
Decays into invisible bosons, $\tau \to \ell a_{\rm inv}$, have been carried out by ARGUS~\cite{ARGUS:1995bjh}, Belle~\cite{Belle:2025bpu} 
and Belle-II~\cite{Belle-II:2022heu} and constrain these BR's to be as low as $\mathcal{O}(10^{-3}$).
When the ALP is short-lived, it leads to tau decays to neutrino-less three-lepton final states, such as $\tau ^+ \to \mu^+ \mu^+ \mu^-$.
We refer to this case as $\tau \to \ell a_{\rm vis}$ bounds.
These are constrained by BaBar~\cite{BaBar:2010axs}, Belle~\cite{Hayasaka:2010np}, ATLAS~\cite{ATLAS:2016jts}, CMS~\cite{CMS:2020kwy}, LHCb~\cite{LHCb:2014kws} and Belle-II~\cite{Belle-II:2024sce}.
In setting the constrains we use the world best limit as collected by PDG~\cite{ParticleDataGroup:2024cfk},
\begin{align}
    &\mathcal{B}(\tau^- \to e^- e^+ e^-) < 2.7 \times 10^{-8},
    \\
    &\mathcal{B}(\tau^- \to \mu^- \mu^+ \mu^-) < 1.9 \times 10^{-8},
    \\
    &\mathcal{B}(\tau^- \to \mu^- e^+ \mu^-) < 1.7 \times 10^{-8},
    \\
    &\mathcal{B}(\tau^- \to e^- \mu^+ \mu^-) < 2.7 \times 10^{-8},
    \\
    &\mathcal{B}(\tau^- \to e^- e^+ \mu^-) < 1.8 \times 10^{-8},
    \\
    &\mathcal{B}(\tau^- \to e^- \mu^+ e^-) < 1.5 \times 10^{-8}.
\end{align}
These are dominated by constraints from BaBar, Belle, and Belle-II, where tau production proceeds via $e^+e^- \to \tau^+\tau^-$.
When setting constraints on leptophilic ALP parameter space we therefore require that the ALP decays within 1 cm, assuming an average energy of
\begin{equation}
    \langle E_a\rangle \sim \frac{\sqrt{s}}{4} \left(1 + \frac{m_a^2}{m_\tau^2}-\frac{m_\ell^2}{m_\tau^2}\right),
\end{equation}
where $m_\ell$ is the mass of the lepton produced in $\tau \to \ell a$ decays.
For $\tau \to \ell a_{\rm inv}$, we require that the ALP does not decay within 3 m.
The excluded regions are shown in light (dark) grey in \cref{fig:LFV_anarchy,fig:LFV_hierarchy} for visible (invisible) ALPs.

\emph{The tau lifetime} is well known in the SM.
The leptonic and hadronic branching ratios of the $\tau$ are known, in some cases, to N$^3$LO \cite{Davier:2005xq,Banerjee:2020rww,Pruna:2017upz}, while the experimental determinations have $\sim 0.1\%$ precision \cite{ParticleDataGroup:2024cfk}.  
The fit of all observed branching ratios is consistent with the SM at the per-mille level~\cite{ParticleDataGroup:2024cfk}. 
We use this to impose a constraint of $\mathcal{B}(\tau \to \ell a) < 10^{-3}$, which is independent of the decay properties of $a$.
This constraint is shown as a dashed black line in \cref{fig:LFV_anarchy,fig:LFV_hierarchy}.

\emph{Flavor-violating muon decays} can also be directly constrained.
Experimental bounds on $g_{e \mu}/f_a$ are derived from $\mu\to e a$ decays, with ALP being either visible or invisible. 
In our plots, $\mu \to e a_{\rm inv}$ bounds are the most relevant, since for such small masses and weak couplings the ALP is almost always long-lived for $m_{a} < m_\mu - m_e$.
We show the experimental limits from TWIST~\cite{TWIST:2014ymv}, PIENU~\cite{PIENU:2020loi}, and Refs.~\cite{Derenzo:1969za,Bilger:1998rp} (limits from Jodidio et al.~\cite{Jodidio:1986mz} are only valid below 12~MeV).
These curves are the analogues of the isotropic case shown in Ref.~\cite{Calibbi:2020jvd}.

\emph{Off-shell ALP exchange} in $\mu \to e \gamma$, $\mu^+ \to e^+ e^+e^-$, and $\mu \to e$ conversion, as well as analogous channels for taus, also constrain leptophilic ALPs.
In the region of interest for this study, these constraints are subdominant relative to direct emission in lepton decays, however, they can dominate for $m_a > m_\tau - m_e$.
For comparison, in our figures we show the limits on $f_a$ from lepton decays to single photons such as $\mu \to e \gamma$, following Ref.~\cite{Bauer:2021mvw}.
Recently, MEG-II~\cite{MEGII:2025gzr} placed a world-leading limit on $\mathcal{B}(\mu^+ \to e^+ \gamma) < 1.5 \times 10^{-13}$ at 90\%~CL, constraining decay constants as large as $f_a \gtrsim \mathcal{O}(10^{5})$~GeV in a LFV anarchical scenario.
This is to be compared with lower limits of $f_a\gtrsim \mathcal{O}(10^{7})$ from tau decays such as $\tau \to e a$.
Tau decays to a lepton and a photon benefit from the larger ALP couplings, but, for flavor-anarchical scenarios, still provide weaker constraints due to the weaker limits on the branching ratios.
Here, we show limits from BaBar, $\mathcal{B}(\tau \to e \gamma) < 3.3\times 10^{-8}$~\cite{BaBar:2009hkt}, and Belle, $\mathcal{B}(\tau \to e \gamma) < 4.2\times 10^{-8}$~\cite{Belle:2021ysv}, both at 90\%~CL.

\emph{Astrophysical constraints} are relevant at larger decay constants (lower couplings) and are most effective for ALP masses below $m_a \sim \mathcal{O}(100)$~MeV.
In particular, the LFC coupling to electrons and muons are strongly constrained by the observation of SN1987A.
There are multiple mechanisms to derive a limit, depending on the imposed physical requirement.
The most well-known is the cooling argument, based on the fact that weakly-coupled ALPs can be produced and escape the star, carrying away energy from deep inside the progenitor.
Given that the observed neutrinos from SN1987A are consistent with the energy loss expected from SM processes within a proto-neutron star, an additional dominant cooling mechanism is ruled out.
We show limits on $g_{ee}/f_a$ and $g_{\mu\mu}/f_a$ based on this criterion directly from \cite{Calibbi:2020jvd} as a lighter shade of grey and in dashed lines.
Supernova cooling limits on the LFV coupling $g_{e\mu}$ from~\cite{Li:2025beu} are also shown in the same style.  
The $g_{e\mu}$ coupling allows for $e-\mu$ coalescence production of ALPs in the proto-neutron star, which is the dominant process for SN cooling constraint on this coupling for ALP masses in the range $115\,\text{MeV}\lesssim m_a \lesssim 280$\,MeV. 
Note that in the figures we plot each of these bounds (labeled as ``supernova $g_{ee,\mu\mu,e\mu}$") individually, where we consider each of the $g_{\ell\ell'}$ couplings is non-zero at a time. 
This is not entirely correct, for each flavor pattern one should recompute the SN cooling bounds taking all the processes into account simultaneously. However, since one of the couplings typically dominates, this suffices for our purposes -- indicating that there is significant open parameter space for leptophilic ALPs. 

Cooling is not the only way to constrain ALPs with supernovae, however.
For instance, Ref.~\cite{Fiorillo:2025sln} derives limits on $g_{ee}/f_a$ by arguing that the energy outflow induced by the propagating but ``trapped" ALPs would stall the supernova explosion.
These constraints are sensitive to the exact details of the supernova model, and can thus be viewed as less conservative (i.e.~the exact shape of the constraints is more dependent on the assumptions). For the leptophilic ALP they are relevant, since they constrain a portion of the parameter space where ALPs are in the so-called ``trapped regime", where ALPs are too strongly coupled to the plasma to escape the star. These constraints are labeled as ``low energy SN ($g_{ee}$)'' in \cref{fig:LFC} and \cref{fig:LFC_tauphilic} (left), as opposed to the previously mentioned cooling bounds which are labeled as ``SN cooling ($g_{ee}$)''. 

\emph{Beam dumps} can pose meaningful constraints on the LFC case despite the smaller ALP coupling to electrons, as shown in \cref{fig:LFC} and \cref{fig:LFC_tauphilic} (left).
In particular, the E137~\cite{Bjorken:1988as}, E141~\cite{Riordan:1987aw}, and Orsay~\cite{Davier:1989wz} electron beam dumps have been long-known to place limits on axion-like particles~\cite{Bjorken:2009mm,Andreas:2010ms}.
For E137, we plot the results of Ref.~\cite{Eberhart:2025lyu} for the flavor universal case, and Ref.~\cite{Liu:2016qwd} and Ref.~\cite{Araki:2021vhy}  for the flavor-conserving muon-phobic ($g_{\mu\mu} =0$) case.
We find that these improve on the limits from CHARM and SINDRUM shown in Ref.~\cite{Altmannshofer:2022ckw}.

\emph{Colliders}, such as  BaBar~\cite{BaBar:2020jma} and Belle~\cite{Belle:2022gbl} have set limits on leptophilic scalars that can be adapted to the ALP scenarios considered here.
The searches are based on $e^+e^- \to \tau^+\tau^- a\, (a \to \ell^+\ell^-)$ and set a lower limit on $f_a$ of $\mathcal{O}(500\,\text{GeV})$ for $m_a > 40$\,MeV.
In most of the excluded parameter space, ALPs are promptly-decaying particles, and therefore, cannot be constrained by traditional beam dumps setups.
In the near future, one can also envisage new constraints on ALPs from the decays of $W^\pm$ bosons at the LHC, produced, for example, in $W\to \tau (a\to \ell^+\ell^-) \nu_\tau$ decays~\cite{Fei:2024qtu}.
These can help constrain short-lived ALPs heavier than the tau up to $m_W - m_\tau$.

\emph{Fixed target experiments} such as NA64 can constrain ALPs appearing either as missing energy~\cite{Gninenko:2719646,NA64:2021xzo}, through their decays to $e^+e^-$~\cite{NA64:2019auh,NA64:2021aiq}, or, in the future, through LFV processes in the target, such as $\mu \to e$~\cite{Gninenko:2022ttd} or $\mu \to \tau$~\cite{Radics:2023tkn}.
Searches have been done using both an electron and a muon beam~\cite{NA64:2024klw}, shown as NA64$e$ or NA64$\mu$, respectively.
For the flavor-conserving scenarios, we draw the contours from Ref.~\cite{Eberhart:2025lyu}.
Because of the higher energy at fixed target experiments and the access to relatively short-lived particles, they can reach ALP masses larger than the tau mass~\cite{Davoudiasl:2021mjy,Radics:2023tkn,Batell:2024cdl,Ponten:2024grp}.

\emph{Anomalous magnetic moments} of the electron and muon are measured very precisely and are also sensitive to novel contributions from an ALP running in the loop.
In the case of the muon, the difference between the experimental world-average for $a_\mu = (g_\mu -2)/{2}$, including the recent measurement by the Muon g-2 experiment~\cite{Muong-2:2025xyk}, and the latest SM theoretical prediction~\cite{Aliberti:2025beg} is given by
\begin{equation}
    \Delta a_\mu = a_\mu^{\rm EXP} - a_\mu^{\rm TH} = (384 \pm 637) \times 10^{-12}.
\end{equation}
Note that the SM theory prediction is based on lattice calculations of the hadronic vacuum polarization contribution~\cite{Borsanyi:2020mff,Bazavov:2024eou,Djukanovic:2024cmq,RBC:2024fic}, while the dispersion based estimates are in tension with it and with the experimental value for $a_\mu$ (see the discussion in~\cite{Aliberti:2025beg}).
Since ALPs contribute to $a_\mu$ with a negative sign to $\Delta a_\mu$, 
we can place a limit on the ALP contribution by requiring $\Delta a_\mu > -890 \times 10^{-12}$ at $2\sigma$.
For the flavor-universal LFC ALP benchmark, this limit excludes decay constants as large as $f_a \gtrsim \mathcal{O}({10^{2}})$~GeV, closing the window between meson decay and beam-dump constraints.
In the case of the electron $\Delta a_e$, the experimental landscape is less conclusive, with the two leading measurements disagreeing at more than $5\sigma$~\cite{Parker:2018vye,Morel:2020dww}.
For that reason, in \cref{fig:LFC}, we only show the limits from $a_\mu$ and not $a_e$.
In deriving the excluded regions we use the expressions in Ref.~\cite{Bauer:2021mvw}, neglecting diagrams proportional to the ALP coupling to two photons that are subdominant in our region of interest.

\section{Conclusions} 
\label{sec:conclusions}

We investigated the phenomenology of leptophilic ALPs, with a particular focus on ALPs that couple only to leptons. 
These can be produced in tau decays, providing a new opportunity to discover or exclude ALPs with couplings that would otherwise be beyond the reach of flavor, beam dump, and collider observables (see, e.g.,~\cite{Bauer:2017ris,Cornella:2019uxs,MartinCamalich:2020dfe,Calibbi:2020jvd,Bauer:2021mvw}). 
We considered both flavor-violating and flavor conserving ALP couplings to leptons, with or without flavor universality, allowing for ALP decays to be prompt, displaced/long-lived, or escape detection entirely --  appearing as missing energy and momentum. 
We derived new bounds and estimated the sensitivity of current and future experiments to the decay constant $f_a$ of the ALP. 
Ultimately, the SHiP experiment will be well poised to probe regions of parameter space with decay constants as high as $\mathcal{O}(10^9\,\text{GeV})$ for flavor-violating ALP scenarions in the mass range $m_\mu + m_e < m_a < m_\tau - m_e$.  

Past searches for long-lived particles at the CHARM and BEBC beam dumps at CERN already constrain the parameter space of leptophilic ALPs, reaching decay constants as high as $f_a \sim \mathcal{O}(10^8\,\text{GeV})$ for the lepton-flavor-violating case.
We have also presented the event-rate sensitivity (corresponding to $90\%$~CL sensitivity in the case of negligible background counts) of currently-operating experiments: MicroBooNE, ICARUS, and NOvA at Fermilab, and protoDUNE and FASER at CERN, as well as of future experiments such as the next-generation forward physics facility at the LHC, FASER-2, the dedicated large-acceptance beam-dump at CERN's SPS, SHiP, and the DUNE near detector.
Eventually, with its full dataset, SHiP is expected to surpass all other projects thanks to the large number of taus produced at the target and the large acceptance of the detector. 
The improvement over FASER and FASER-2 in the short-lived regime, where a large number of ALPs are produced but only a few travel far enough to decay inside the detector, is small.
However, in the long-lived regime, where a small number of ALPs are produced but only a few decay as they pass through the detector, the improvement over other projects can be significant, with over an order of magnitude improvement over CHARM and BEBC constraints.

For flavor-violating scenarios with a hierarchy between flavor-violating and flavor-conserving couplings, $g_{\ell_i \ell_j} \ll g_{\ell_i \ell_i}$, the CHARM and BEBC exclusions are closer to $f_a \sim \mathcal{O}(10^6 - 10^7\,\text{GeV})$, but they do constrain new parameter space.
If flavor-violating couplings are heavily suppressed, keeping the ALP production rates from $\tau \to \ell a$ decays sizable requires smaller values of $f_a$. 
In this limit, the ALP becomes very short-lived above the dimuon threshold, quickly degrading the sensitivity of conventional beam dumps.
In this scenario, FASER-2 and SHiP have an advantage since FASER-2 will be capable of producing ALP with larger boosts and SHiP will be located closer to the target.

In the case of lepton-flavor-conserving couplings, ALP production can only take place via 3-body and 4-body $G_F$-suppressed decays, and thus the experimental reach is much weaker.
For a flavor universal scenario, $g_{ee}=g_{\mu \mu}=g_{\tau \tau}$, we find that FASER-2 and SHiP will be able to probe new parameter space for $f_{a} \sim \mathcal{O}(10 - 10^4)\,\text{GeV}$.
For such low values of $f_a$, ALPs are only long-lived, if they are lighter than $m_a < 2m_\mu$, restricting the search strategies to just the decays of  ALPs to $e^+e^-$ pairs.
This conclusion can change, if the flavor-conserving couplings are non-universal.
In particular, if $g_{\mu \mu} \ll g_{ee}, g_{\tau \tau}$, then the partial decay  width for $a\to\mu^+\mu^-$ decays can be as suppressed as the partial decay width for $a\to e^+e^-$ (or even further suppressed), extending the experimental reach of beam dump experiments to higher ALP masses, above the di-muon threshold.

We conclude by emphasizing that there still remain many opportunities to probe physics beyond-the-Standard Model with tau leptons.
We have presented one example based on long-lived particle production, the leptophilic ALP, but one may equally consider the production of other exotic short-lived states.
In that case, the new-physics searches are restricted to less-intense, but cleaner environments such as $e^+e^-$ colliders.
Given the large numbers of $\tau$ leptons set to be produced at modern colliders, one can expect significant progress from Belle-II~\cite{Belle-II:2018jsg}, LHCb, ATLAS, and CMS, and eventually at a Super Tau-Charm factory~\cite{Achasov:2023gey}. 
We plan to present a more detailed study of exotic signals due to short-lived dark particles produced in tau decays in a future work~\cite{Ema:inprep}.

\begin{acknowledgments}
We thank Brian Batell, Simon Knapen, Stephen Mrenna, Maksym Ovchynnikov for helpful discussions.  The work of MH is supported by the Neutrino Theory Network Program Grant \#DE-AC02-07CHI11359 and the US DOE Award \#DE-SC0020250. AR acknowledge support from the National Science Foundation (Grant No. PHY-2020275) and to the Heising-Simons Foundation (Grant No. 2017-228). JZ and TM acknowledges support in part by the DOE grant de-sc0011784, and NSF grants OAC-2103889, OAC-2411215, and OAC- 2417682.  Fermilab is managed by FermiForward Discovery Group, LLC, acting under Contract No.89243024CSC000002 with the U.S. Department of Energy, Office of Science, Office of High Energy Physics. This research was supported by the Munich Institute for Astro-Particle and BioPhysics (MIAPP) which is funded by the Deutsche Forschungsgemeinschaft (DFG, German Research Foundation) under Germany's Excellence Strategy -- EXC-2094 -- 390783311.  Part of this research was performed at the Aspen Center for Physics, which is supported by National Science Foundation grant PHY-1607611.

Our results can be reproduced with a publicly available code on \gitlink under CC-BY-NC-SA.

\end{acknowledgments}

\appendix

\section{Impact of tau production uncertainties}
\label{app:tauprod}

The total number of charmed mesons and their energy spectrum at fixed target experiments is subject to sizeable uncertainties.
For the $\tau$ yield, the uncertainty on $D_s^\pm$ production is the most important one ($D^\pm$ have a smaller branching ratio to taus). 
In this appendix, we discuss how these uncertainties can impact the experimental sensitivities to leptophilic ALPs from tau decays, using NOvA and SHiP for illustration.
We compare the methodology of \cref{sec:alpflux}, based entirely on our \pythia simulation, with an approach based on a simplified parameterization of the $D_{s}^\pm$ production rate and kinematics, following Ref.~\cite{Schubert:2024hpm}.

A standard parameterization for $D^\pm$ and $D^0$ production cross section at fixed target experiments is~\cite{Lourenco:2006vw}
\begin{equation}
    \frac{d \sigma}{d p_T dx_F} \propto (1 - |x_F|)^n e^{- a p_T - b p_T^2},
\end{equation}
where $p_T$ is the transverse momentum of the outgoing $D$ meson and $x_F = p_{z}^{\rm COM}/p_{z{-\rm max}}^{\rm COM}$ is the Feynman-$x$ calculated in the center of mass of the $pp$ collision.
In this procedure, the normalization is fixed by the charm production cross section.
The number of $\tau$ leptons $N_\tau$ produced in $pp$ collisions is approximately given by
\begin{equation}\label{eq:normalization}
    N_{\tau} = N_\text{pp-col} \frac{\sigma_{c\overline c}}{\sigma_{pp}^{\rm inelastic}} f_{c \to D_{(s)}} \mathcal{B}(D_s \to \tau),
\end{equation}
where $N_\text{pp-col}$ is the number of proton-proton collisions, $\sigma_{c\bar c}$ the total charm-pair production cross-section, and  $\sigma_{pp}^{\rm inelastic}$ the total inelastic $pp$ cross section.
Note that in the above expression we neglected the $D^\pm$ contribution.
The fragmentation fraction is given by $f_{c \to D_s} = 0.079$~\cite{Gladilin:2014tba} and the branching ratio of all the modes containing $\tau$ set to $\mathcal{B}(D_s\to \tau) = 5.4\%$ as in the main text.

In Ref.~\cite{Schubert:2024hpm}, the parameters $n$, $a$, $b$ and as well as the charm production cross section $\sigma_{c\bar c}$ were tuned simultaneously to a broad set of results generated from a custom version of \pythia that incorporates \textsc{FONLL} matrix elements \cite{Cacciari:1998it,Cacciari:2001td} and used \textsc{LHAPDF6}~\cite{Buckley:2014ana} and the \textsc{NNPDF4.0} set of parton-distribution functions~\cite{NNPDF:2021njg}.

\begin{figure*}[t]
    \centering
    \includegraphics[width=0.49\textwidth]{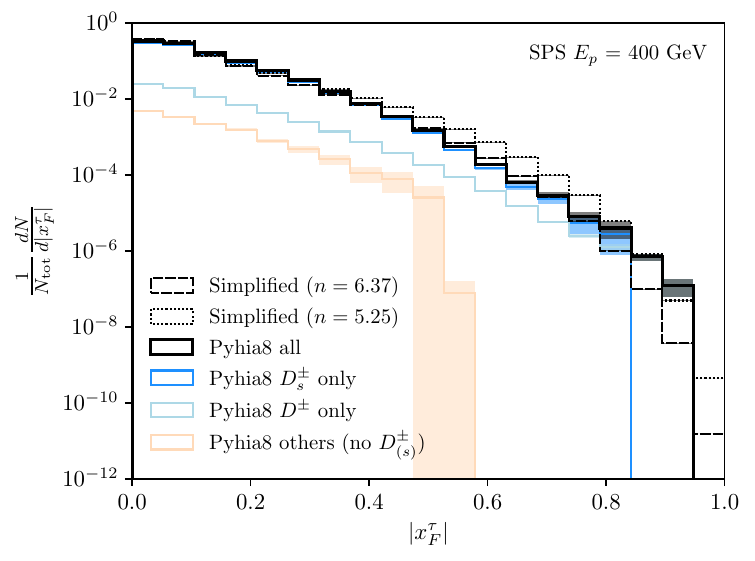}
    \includegraphics[width=0.49\textwidth]{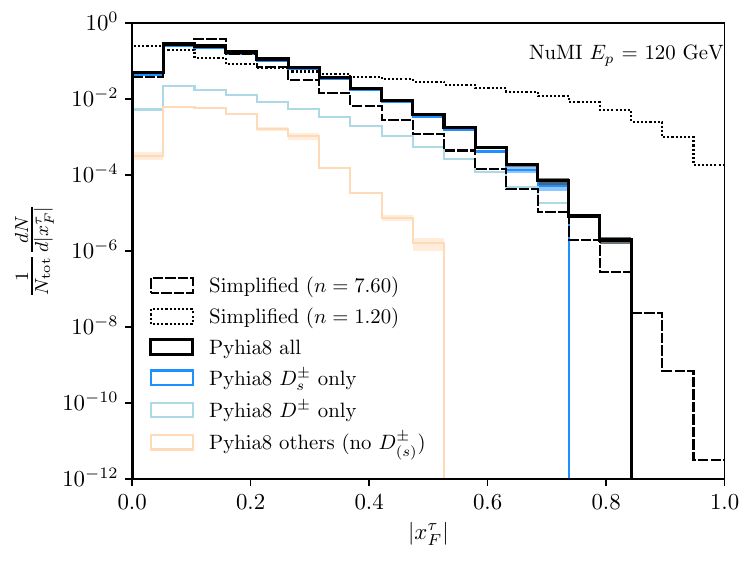}
    \\
    \includegraphics[width=0.49\textwidth]{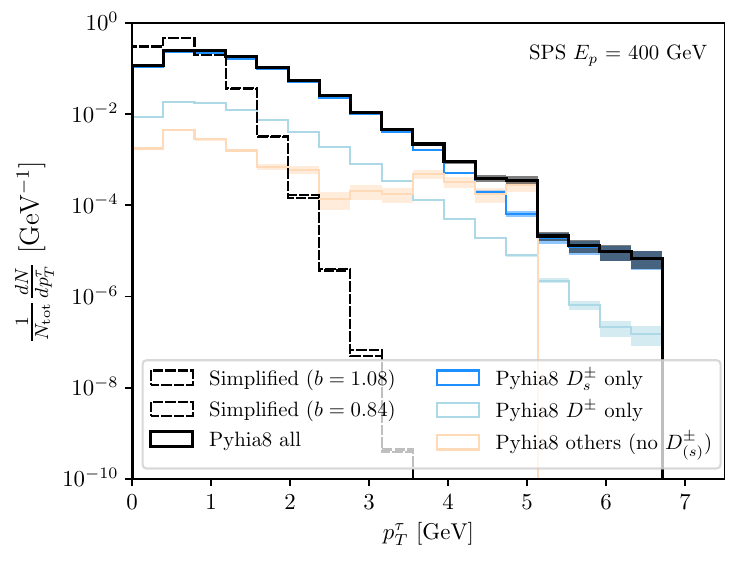}
    \includegraphics[width=0.49\textwidth]{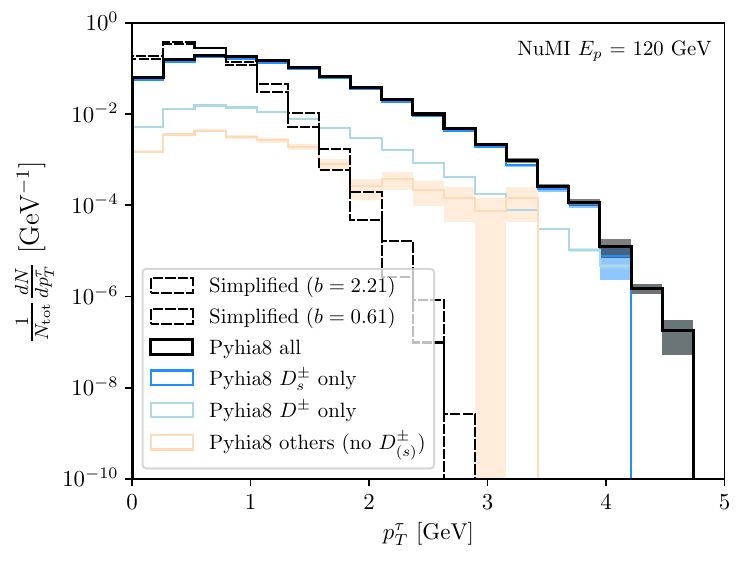}
    \caption{The $x_F$ (top row) and $p_T$ (bottom row) distributions of $\tau^\pm$ leptons produced in primary $pp$ collisions at the NuMI and SPS target in our \pythia simulation (solid black lines) compared with the simplified treatment following the parameterization in~\cite{Schubert:2024hpm} (dashed grey lines).
    In this simplified treatment, all taus are produced by $D_s^\pm$, which are assumed to follow the same kinematics as the $D^\pm$ of \cite{Schubert:2024hpm}.
    We also show our \pythia results including taus only from $D_s^\pm$ mesons (blue solid), $D^\pm$ mesons (light blue solid), and all other sources of taus (cream solid).
    \label{fig:xFandpT}
    }
\end{figure*}

Adopting the prescription of Ref.~\cite{Schubert:2024hpm}, we vary these inputs in concert and propagate the resulting spread to the projected ALP coverage.
While this procedure is not a rigorous estimate of the theoretical uncertainty, it illustrates how variations in the charm-production distributions can impact the experimental reach of searches for leptophilic ALPs produced from tau decays.
In the absence of a dedicated study for $D_s^\pm$, we adopt the parameters and ranges quoted for $D^\pm$.
In what follows, we quote the range of values used in \cite{Schubert:2024hpm}:
\begin{align}
    \text{NuMI:}& \,\, (n, b) = (4.4 0\pm1.60, \,1.41\pm 0.40),
    \\
    \text{SPS:}& \,\, (n, b) = (5.81\pm1.3, \,0.96\pm 0.06),
\end{align}
where we also set $a = 0$. 
We find that the number of taus per POT with our \pythia simulation is in agreement with the normalization of \cite{Schubert:2024hpm} using \cref{eq:normalization}.
In our comparisons below we normalize the simplified simulations according to the results of the fit to $\sigma_{c\bar c}$.

\begin{figure}
    \centering
    \includegraphics[width=0.49\textwidth]{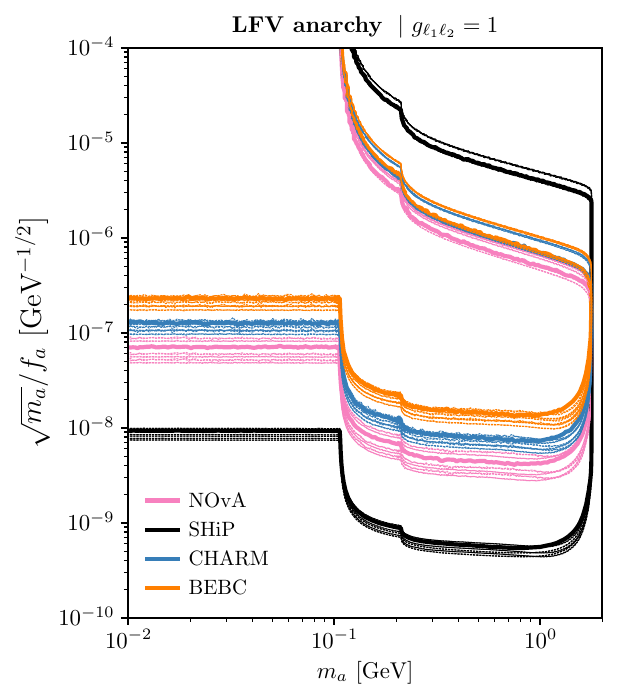}    
    \caption{
    The projected event-rate sensitivities of NOvA (pink), CHARM (blue), BEBC (orange), and SHiP (black) using our standard \pythia simulation (thick solid line) and the simplified simulation based on a parameterization of $D_s^\pm$ production with the $2\sigma$ ranges of input normalization and shape parameters from~\cite{Schubert:2024hpm} (thin dashed lines).
    While the normalization uncertainty impacts mostly the upper limit (small $f_a^{-1}$), the difference between the \pythia results and this simplified simulation are also noticeable on the lower limit (large $f_a^{-1}$) due to the uncertainties on the kinematics of the $D_s^\pm$ produced and the resulting boost factor for the ALPs.
    }
    \label{fig:uncertainty_on_limits}
\end{figure}

In \cref{fig:xFandpT}, we compare the resulting $p_T$ and $x_F$ distribution of taus from our \pythia simulations with those obtained from the simplified procedure outlined above.
We plot the results of the simplified approach using the extremes of the $2\sigma$ allowed regions for the parameters in \cite{Schubert:2024hpm}.
Where not specified, we use $n = 4.4$ and $b=1.41$ for NuMI and $n = 5.81$ and $b = 0.96$ for SPS.
While we find some agreement in the longitudinal variable $x_F$, we see a significant deviation in the $p_T$ spectrum with respect to \pythia results.
The $p_T$ distribution directly affects the geometrical acceptance of different experiments to ALPs, and will thus be a source of uncertainty.
The tails of the longitudinal component of the momentum can also extend to higher values in the simplified approach, impacting the largest boost achievable for ALPs and, therefore, the sensitivity of some experiments to the shorter-lived regime of the ALP parameter space.

\Cref{fig:uncertainty_on_limits} compares the resulting event-rate sensitivities of our \pythia results with the simplified approach described above (using the $2\sigma$ extreme values) in the parameter space of ($\sqrt{m_a}/f_a$) versus $m_a$ as some parts of the sensitivity to $\sqrt{m_a}/f_a$ are flat in $m_a$, as opposed to the sensitivity to $1/f_a$.
We see that while the normalization impacts mostly the upper limit, the \pythia simulation and the simplified approach also disagree in the short-lived regime (lower limit) due to the different kinematics of the ALPs.

In the future, perturbative QCD calculations of the double-differential production rates of $D_s$ mesons at the NuMI and SPS targets could refine the $\tau$ production calculation presented here, similarly to what was performed for for SHiP in Ref.~\cite{Bai:2018xum,Maciula:2019clb} and for FASER in Refs.~\cite{Bai:2020ukz,Bai:2022xad,Bhattacharya:2023zei,Buonocore:2023kna}.

\section{Parametrized tau production}
\label{app:analytical_fits}

In the main text, we describe our simulation using \pythia events.
In this appendix, we present an alternative method for fast simulation of tau production at NuMI, SPS, and at the LHC. 
We do not rely on this method in our results, but include it here for future reference.
In analogy to parameterizations historically used for charmed hadron production, we can describe the kinematics of the \emph{tau lepton} in terms of its transverse momentum $p_T$ and $x_F = p_{z}^{\rm COM}/p_{z{-\rm max}}^{\rm COM}$, the Feynman-$x$ of the tau calculated in the center of mass of the $pp$ collision.
We find that, up to an overall normalization factor, a satisfactory parameterization of the tau kinematics at the three experiments can be found with the following exponential functions:
\begin{align}
    \frac{dN_\tau}{dx_F dp_T} & = \left[\sum_{i=1}^3 g_i \Gamma(p_T, \alpha_i, \lambda_i) \right] \left[\sum_{j=1}^3 c_j\exp({-a_j|x_F|^{n_j}}) \right],
\end{align}
where $\Gamma(x, \alpha, \lambda) = x^{\alpha -1 } e^{-\lambda x} \lambda^\alpha/\Gamma(\alpha)$ is the gamma distribution and $\Gamma(\alpha)$ the gamma function.
Here $c_i$, $g_i$, $n_i$, $\alpha_i$, and $\lambda_i$ are all constants that we fit to the output of our \pythia simulations.
As expected, we find that $x_F$ and $p_T$ are, to a good approximation, uncorrelated, so our fits can be done separately in terms of the two one-dimensional distributions.

\begin{figure*}[t]
    \centering
    \includegraphics[width=0.49\textwidth]{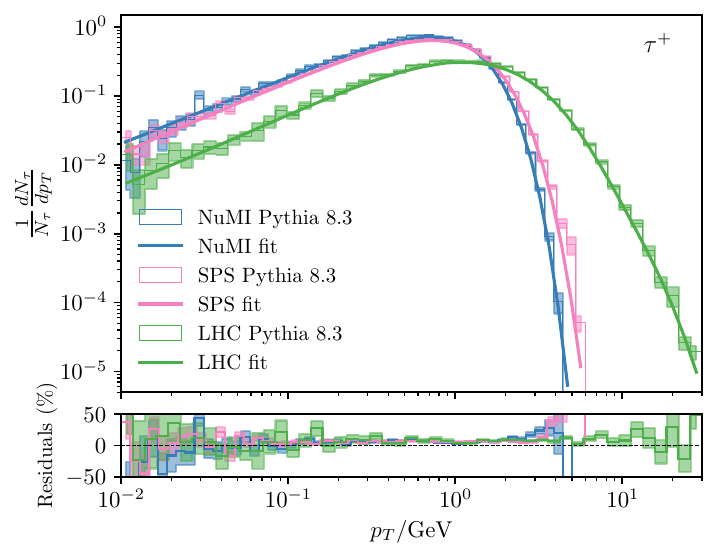}
    \includegraphics[width=0.49\textwidth]{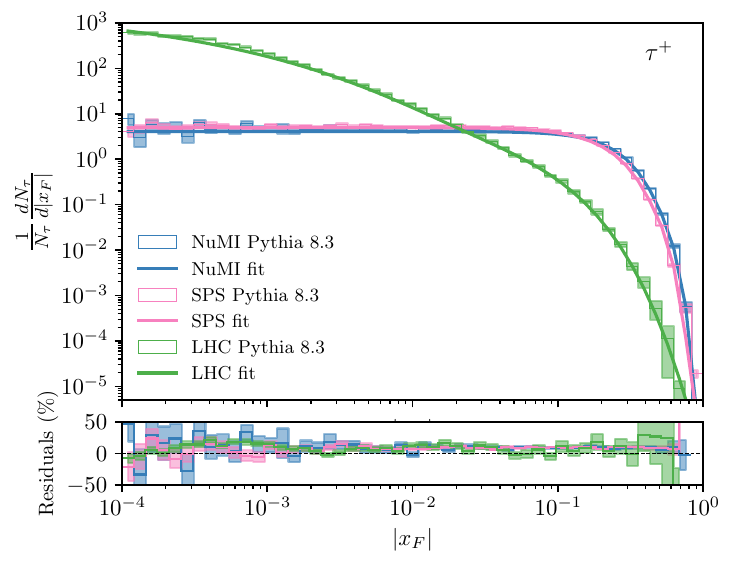}
    \includegraphics[width=0.49\textwidth]{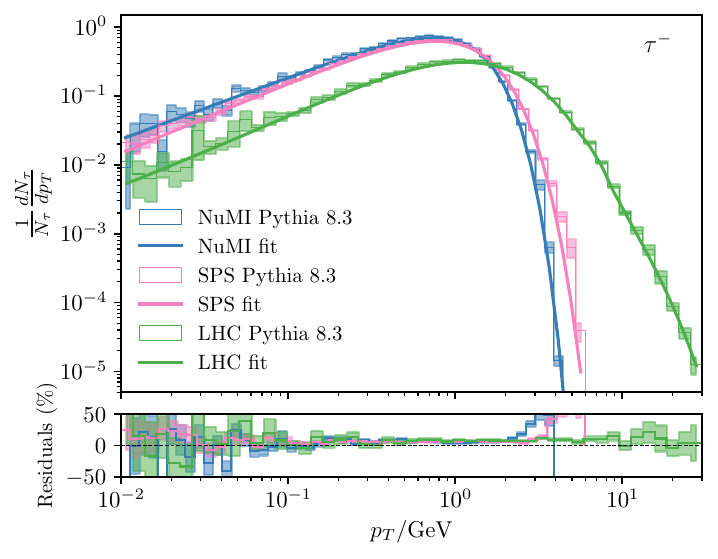}
    \includegraphics[width=0.49\textwidth]{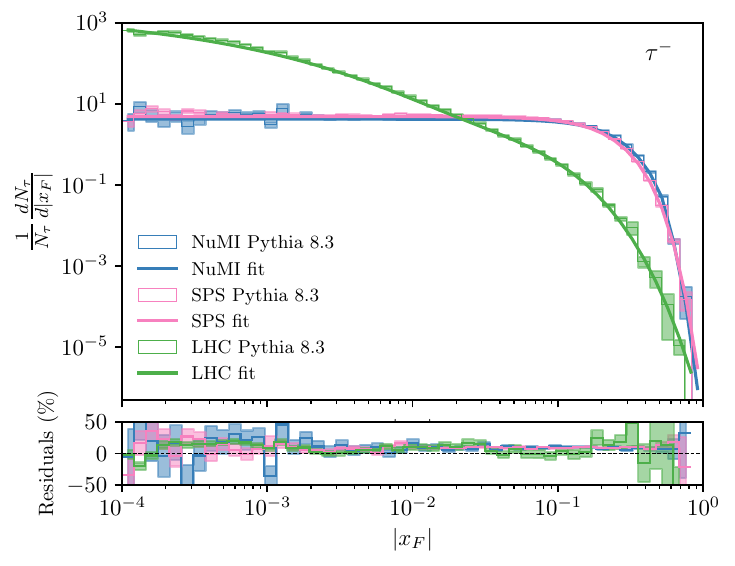}
    \caption{The $|p_T|$ (left) and $|x_F|$ (right) distributions of $\tau^+$ mesons from primary $pp$ collisions at the NuMI and SPS targets as well as from the LHC.
    The results from our \pythia simulations (solid histograms) are fit using analytic functions (see text for details).
    The best fit curves are shown as solid lines.
    The bottom panel shows the relative residuals between simulation and our fit.
    }
    \label{fig:xFandpT_custom}
\end{figure*}

The resulting best-fit curves are shown alongside the \pythia simulation output in \cref{fig:xFandpT_custom} and the best-fit parameters are listed in \cref{tab:fitparams_combined}.
In $x_F$ ($p_T$), we restrict the range of the fit to $|x_F| < 1$, and $p_T < 5$~GeV) for NuMI, $p_T < 6$~GeV for SPS, and $p_T < 30$~GeV for the LHC.
With $x_F$ and $p_T$ distributions fixed in the center of mass of the $pp$ collision, one can sample the remaining azimuthal angular variable in an uniform way and boost the system to the appropriate lab frame.

\begin{table*}[t]
\centering
\small
\renewcommand{\arraystretch}{1.3}
\begin{tabular}{|c|ccc|ccc|}
\hline
\hline
\multirow{2}{*}{\textbf{Parameter}} & \textbf{NuMI} & \textbf{SPS} & \textbf{LHC} & \textbf{NuMI} & \textbf{SPS} & \textbf{LHC} 
\\
\cline{2-7}
  & \multicolumn{3}{c|}{$\tau^+$} & \multicolumn{3}{c|}{$\tau^-$} 
\\
\hline
\hline
\multicolumn{7}{|c|}{\textbf{Gamma distribution fits in $p_T$}} \\
\hline
$g_1$         & 2047052.33 & 2869884.82 & 29.95      & 84411.01    & 9496674.67 & 22.30 \\
$m_1$         & 1.684      & 1.440      & 0.857      & 1.896       & 1.511      & 0.850 \\
$\mu_1$       & 1.578      & 1.729      & 2.682      & 1.473       & 1.674      & 2.839 \\
$\lambda_1$   & 0.964      & 0.875      & 0.660      & 1.219       & 0.990      & 0.627 \\
\hline
$g_2$         & 0.570      & 0.570      & 4.515      & 0.570       & 0.570      & 3.256 \\
$m_2$         & 1.000      & 1.000      & 0.695      & 1.000       & 1.000      & 0.588 \\
$\mu_2$       & 89.18      & 89.18      & 2.105      & 89.18       & 89.18      & 2.555 \\
$\lambda_2$   & 84.40      & 84.40      & 0.838      & 84.40       & 84.40      & 0.535 \\
\hline
$g_3$         & 0.570      & 0.570      & 0.109      & 0.570       & 0.570      & 4.628 \\
$m_3$         & 1.000      & 1.000      & 6.769      & 1.000       & 1.000      & 2.166 \\
$\mu_3$       & 89.18      & 89.18      & 116.25     & 89.18       & 89.18      & 99.912 \\
$\lambda_3$   & 84.40      & 84.40      & 0.00347    & 84.40       & 84.40      & 0.662 \\
\hline
\hline
\multicolumn{7}{|c|}{\textbf{Exponential fits in $x_F$}} \\
\hline
$r_1$         & 48.69      & 0.339      & 263.92     & 5112.29     & 3810.67    & 351.77 \\
$a_1$         & 16.95      & 31519936.65 & 17.36     & 18.32       & 16.70      & 17.10 \\
$n_1$         & 4.559      & 35.98      & 0.808      & 2.026       & 1.795      & 0.777 \\
\hline
$r_2$         & 3593.68    & 189.77     & 132431.87  & 82.01       & 137.60     & 126552.23 \\
$a_2$         & 18.39      & 17.51      & 23.68      & 119.43      & 2254.07    & 26.02 \\
$n_2$         & 2.090      & 1.807      & 0.321      & 6.873       & 4.767      & 0.342 \\
\hline
\hline
\end{tabular}
\caption{Best-fit parameters for $\tau^+$ and $\tau^-$ production in $pp$ collisions, grouped by transverse momentum ($p_T$) and Feynman-$x$ ($x_F$) distributions. The fits consist of a sum of three gamma distributions in $p_T$ and two exponential terms in $x_F$. Sets NuMI, SPS, and LHC correspond to different collider setups.}
\label{tab:fitparams_combined}
\end{table*}

\section{Details on ALP production and decay}
\label{app:details}

In this appendix, we summarize the computational details of the ALP production and decay rates.

\subsection{The $\tau \to \pi \nu_\tau a$ decay}

We begin with the decay mode $\tau^- \to \pi^- \nu_\tau a$.
The amplitude of this process is given by
\begin{align}
	i\mathcal{M}
	&= 
	\begin{tikzpicture}[baseline=(a)]
	\begin{feynman}[inline = (base.a), horizontal=a to c]
		\vertex [label=\({\scriptstyle \tau, P}\)] (a);
		\vertex [right = 1.5 of a] (b);
		\vertex [below right = of b, label=0:\({\scriptstyle \nu_\tau, p_2}\)] (c);
		\vertex [above right = of b, label=0:\({\scriptstyle \pi, p_1}\)] (d);
		\vertex [right = 0.75 of a] (v1);
		\vertex [above right = of v1, label=180:\({\scriptstyle a, p_3}\)] (e);
		\diagram*{
		(a) -- [fermion] (v1) -- [fermion] (b) -- [fermion] (c),
		(d) -- [scalar] (b),
		(v1) -- [scalar] (e),
		};
	\end{feynman}
	\end{tikzpicture}
	+
	\begin{tikzpicture}[baseline=(a)]
	\begin{feynman}[inline = (base.a), horizontal=a to c]
		\vertex [label=\({\scriptstyle \tau, P}\)] (a);
		\vertex [right = of a] (b);
		\vertex [below right = 1.5 of b, label=0:\({\scriptstyle \nu_\tau, p_2}\)] (c);
		\vertex [above right = of b, label=0:\({\scriptstyle \pi, p_1}\)] (d);
		\vertex [below right = 0.75 of b] (v1);
		\vertex [above right = of v1, label=0:\({\scriptstyle a, p_3}\)] (e);
		\diagram*{
		(a) -- [fermion] (b) -- [fermion] (v1) -- [fermion] (c),
		(d) -- [scalar] (b),
		(v1) -- [scalar] (e),
		};
	\end{feynman}
	\end{tikzpicture}
	\nonumber \\
	&= -\frac{\sqrt{2} iG_F V_{ud} f_\pi m_\tau g_{\tau\tau}}{f_a(m_\tau^2 - \bar{m}_\tau^2)}
	\bar{u}_\nu P_R \left[m_\tau \slashed{p}_{12}-\bar{m}_{\tau}^2\right]u_\tau,
\end{align}
where $P_R = (1+\gamma_5)/2$, $p_{ij} = p_i + p_j$, and $\bar{m}_\tau^2 = p_{12}^2$ is the invariant mass
of the intermediate off-shell  $\tau$ (in the first diagram).
We need to include the diagram with the ALP attached to the neutrino leg to be consistent with the EW gauge symmetry;
only after this, we obtain the amplitude proportional to $m_\ell$.

The differential decay rate is given by
\begin{align}
	d\Gamma(\tau \to \pi \nu_\tau a) &= \frac{\abs{\mathcal{M}}^2}{2m_\tau}d\Phi_3,
\end{align}
where $d\Phi_3$ is the three-body phase space element 
(with the $(2\pi)^4$ included in the delta function that ensures the total four-momentum conservation).
Averaging over the spin of the initial tau, gives
\begin{align}
	d\Gamma(\tau \to \pi \nu_\tau a) &= \frac{\abs{\bar{\mathcal{M}}}^2}{256 \pi^3 m_\tau^3}d\bar{m}_{\tau}^2 dm_{23}^2,
\end{align}
where $\abs{\bar{\mathcal{M}}}^2$ is the spin averaged amplitude squared and $m_{23}^2 = p_{23}^2$.
Integrating this expression over $m_{23}^2$ from $(m_{23}^2)_+$ to $(m_{23}^2)_-$, where
\begin{align}
	(m_{23}^2)_\pm = (E_2^* + E_3^*)^2 - \left(E_2^*\pm \sqrt{E_3^{*2}-m_a^2}\right)^2,
\end{align}
and 
\begin{align}
	E_2^* = \frac{\bar{m}_{\tau}^2 - m_\pi^2}{2\bar{m}_{\tau}},
	\quad
	E_3^* = \frac{m_{\tau}^2 - \bar{m}_{\tau}^2 - m_a^2}{2\bar{m}_{\tau}},
\end{align}
we obtain~\cref{eq:dGamma_tau2pinua} in the main text.

\subsection{The $\tau \to \rho \nu_\tau a$ decay}

The calculation of partial decay width for $\tau^- \to \rho^- \nu_\tau a$, is analogous to the one for $\tau^- \to \pi^- \nu_\tau a$.
The amplitude 
 is given by
\begin{align}
	i\mathcal{M}
	&= 
	\begin{tikzpicture}[baseline=(a)]
	\begin{feynman}[inline = (base.a), horizontal=a to c]
		\vertex [label=\({\scriptstyle \tau, P}\)] (a);
		\vertex [right = 1.5 of a] (b);
		\vertex [below right = of b, label=0:\({\scriptstyle \nu_\tau, p_2}\)] (c);
		\vertex [above right = of b, label=0:\({\scriptstyle \rho, p_1}\)] (d);
		\vertex [right = 0.75 of a] (v1);
		\vertex [above right = of v1, label=180:\({\scriptstyle a, p_3}\)] (e);
		\diagram*{
		(a) -- [fermion] (v1) -- [fermion] (b) -- [fermion] (c),
		(d) -- [photon] (b),
		(v1) -- [scalar] (e),
		};
	\end{feynman}
	\end{tikzpicture}
	+
	\begin{tikzpicture}[baseline=(a)]
	\begin{feynman}[inline = (base.a), horizontal=a to c]
		\vertex [label=\({\scriptstyle \tau, P}\)] (a);
		\vertex [right = of a] (b);
		\vertex [below right = 1.5 of b, label=0:\({\scriptstyle \nu_\tau, p_2}\)] (c);
		\vertex [above right = of b, label=0:\({\scriptstyle \rho, p_1}\)] (d);
		\vertex [below right = 0.75 of b] (v1);
		\vertex [above right = of v1, label=0:\({\scriptstyle a, p_3}\)] (e);
		\diagram*{
		(a) -- [fermion] (b) -- [fermion] (v1) -- [fermion] (c),
		(d) -- [photon] (b),
		(v1) -- [scalar] (e),
		};
	\end{feynman}
	\end{tikzpicture}
	\nonumber \\
	&= -\frac{\sqrt{2}G_F V_{ud} m_\rho m_\tau f_\rho g_{\tau\tau}}{ f_a (m_{\tau}^2 - \bar{m}_\tau^2)}
	\bar{u}_\nu P_R\slashed{\epsilon}^*\left[\slashed{p}_{12}-m_\tau\right]u_\tau,
\end{align}
where we again note the contribution from the second diagram.
After performing the phase space integrals, averaging over the tau spin, and summing over final state particle polarizations, we obtain~\cref{eq:dGamma_tau2rhonua}.

\subsection{The $\tau \to \ell \nu_\ell \nu_\tau a$ decay}

The calculation of the partial decay width for $\tau^- \to \ell^- \bar{\nu}_\ell \nu_\tau a$, is more complicated, 
since it involves integrating over the four-body  phase space.
The amplitude is given by
\begin{align}
	i\mathcal{M} &= 
	\begin{tikzpicture}[baseline=(a)]
	\begin{feynman}[inline = (base.a), horizontal=a to c]
		\vertex [label=\({\scriptstyle \tau, P}\)] (P);
		\vertex [right = of P] (v1);
		\vertex [above right = of v1, label=180:\({\scriptstyle a, p_1}\)] (p1);
		\vertex [right = of v1] (v2);
		\vertex [above right = of v2, label=0:\({\scriptstyle l, p_2}\)] (p2);
		\vertex [right = of v2, label=0:\({\scriptstyle \nu_\tau, p_3}\)] (p3);
		\vertex [below right = of v2, label=0:\({\scriptstyle \bar{\nu}_l, p_4}\)] (p4);
		\diagram*{
		(P) -- [fermion] (v1) -- [fermion, momentum'=\({\scriptstyle p_{234}}\)] (v2) -- [fermion] (p3),
		(p4) -- [fermion] (v2) -- [fermion] (p2),
		(v1) -- [scalar] (p1),
		};
	\end{feynman}
	\end{tikzpicture}
	+ \cdots
	\nonumber \\
	&= -\frac{2\sqrt{2} G_F g_{\tau\tau} m_\tau}{f_a(p_{234}^2 - m_\tau^2)}
	\left[\bar{u}_\nu P_R \gamma_\alpha (\slashed{p}_{234}-m_\tau) u_\tau\right]
	\left[\bar{u}_l \gamma^\alpha P_Lv_\nu \right],	
\end{align}
where $P_L = (1-\gamma_5)/2$, $p_{234} = p_2 + p_3 + p_4$,
and for brevity we show only the diagram with the ALP attached to $\tau$ and not the one with ALP attaching to the $\nu_\tau$ line.
In the calculation we do ignore the diagrams with the ALP attached to $\ell$ and $\bar{\nu}_l$ since those are suppressed by $m_\ell/m_\tau$
at the amplitude level.
The decay rate is given by
\begin{align}
	d\Gamma = \frac{1}{2m_\tau} d\Phi_4 \left\vert \bar{\mathcal{M}}\right\vert^2,
\end{align}
where $d\Phi_4$ is the differential four-body phase space element.
The spin averaged amplitude squared is given by
\begin{align}
	\left\vert \bar{\mathcal{M}}\right\vert^2 &= \frac{64 G_F^2 \abs{g_{\tau\tau}}^2 m_\tau^2}{f_a^2 (m_\tau^2 - p_{234}^2)^2}
	(p_2 \cdot p_3) p_4^\alpha 
	\nonumber \\
	&\times \left[(m_\tau^2 - p_{234}^2)P_\alpha + 2p_{234\alpha}(P\cdot p_{234} - m_\tau^2)\right].
\end{align}
It depends on $p_2$, $p_3$ and $p_4$ separately only through the form $(p_2 \cdot p_3) p_4^\alpha$.
It is then useful to decompose the four-body phase space integral sequentially as
\begin{align}
	\int d\Phi_4 &= \int_{m_l^2}^{(m_\tau - m_a)^2} \frac{d \bar{m}_\tau^2}{2\pi}
	\int d\Phi_{1;234} 
	\nonumber \\
	&\times \int_0^{(\bar{m}_\tau - m_l)^2} \frac{d \bar{m}_{\nu\nu}^2}{2\pi}
	\int d\Phi_{2;34} \int d\Phi_{3;4}
\end{align}
where $\bar{m}_\tau^2 = p_{234}^2$ and $\bar{m}_{\nu\nu}^2 = p_{34}^2$ 
are the off-shell invariant mass of $\tau$ and $\bar{\nu}_\ell \nu_\tau$ pair, respectively, and
\begin{align}
	\int d\Phi_{1;234} &= \frac{1}{4\pi^2}\int \frac{d^3 p_1}{2E_1} \frac{d^3p_{234}}{2\bar{E}_{234}}
	\delta^{(4)}(P-p_1 - p_{234}),
\end{align}
\begin{align}
	\int d\Phi_{2;34} &= \frac{1}{4\pi^2}\int \frac{d^3 p_2}{2E_2} \frac{d^3p_{34}}{2\bar{E}_{34}}
	\delta^{(4)}(p_{234}-p_2 - p_{34}),
\end{align}
\begin{align}
	\int d\Phi_{3;4} &= \frac{1}{4\pi^2}\int \frac{d^3 p_3}{2p_3} \frac{d^3p_{4}}{2p_4}
	\delta^{(4)}(p_{34}-p_3 - p_{4}),
\end{align}
with $\bar{E}_{234}^2 = \vert \vec{p}_{234}\vert^2 + \bar{m}_\tau^2$ and $\bar{E}_{34}^2 = \vert \vec{p}_{34}\vert^2 + \bar{m}_{\nu\nu}^2$.
We can show, \emph{e.g.}, by going to the $p_{34}/p_{234}$-rest frame and using the Lorentz invariance, that
\begin{align}
	\int d\Phi_{3;4} p_3^\alpha p_4^\beta = \frac{1}{96\pi}\left(\eta^{\alpha\beta} \bar{m}_{\nu\nu}^2 + 2p_{34}^\alpha p_{34}^\beta\right),
\end{align}
and therefore
\begin{align}
	\int d\Phi_{2;34} \int d\Phi_{3;4} (p_2\cdot p_3)p_4^\alpha
	= \frac{I(\bar{m}_\tau, \bar{m}_{\nu\nu})}{1536 \pi^2}\bar{m}_\tau^2 p_{234}^\alpha,
\end{align}
where the function $I(\bar{m}_\tau,\bar{m}_{\nu\nu})$ is defined in~\cref{eq:I_def}.
By using
\begin{align}
	P\cdot p_{234} = \frac{1}{2}\left[m_\tau^2 + \bar{m}_\tau^2 - m_a^2\right],
\end{align}
and performing the angular integral of the ``$1;234$'' system,
we obtain~\cref{eq:dGamma_tau2lnunua}.

\subsection{The $a \to \gamma \gamma$ decay}

Finally, we consider the ALP decay mode $a\to \gamma\gamma$ through the loop of a heavy lepton $\ell$.
This process is identical to the chiral anomaly in the limit $m_\ell \to 0$, 
and therefore we need some care on the regularization. The amplitude is diagrammatically expressed as
\begin{align}
	i\mathcal{M} 
	&= 
	\begin{tikzpicture}[baseline=(a)]
	\begin{feynman}[inline = (base.a), horizontal=a to c]
		\vertex [label=\({\scriptstyle a, P}\)](a);
		\vertex [right = 0.75 of a] (v1);
		\vertex [above right = of v1] (v2);
		\vertex [below right = of v1] (v3);
		\vertex [right = 0.75 of v2,label=0:\({\scriptstyle \gamma, p_1}\)] (p1);
		\vertex [right = 0.75 of v3,label=0:\({\scriptstyle \gamma, p_2}\)] (p2);
		\diagram*{
		(a) -- [scalar] (v1) -- [fermion] (v2) -- [fermion] (v3) -- [fermion] (v1),
		(v2) -- [photon] (p1),
		(v3) -- [photon] (p2),
		};
	\end{feynman}
	\end{tikzpicture}
	+
	\begin{tikzpicture}[baseline=(a)]
	\begin{feynman}[inline = (base.a), horizontal=a to c]
		\vertex (a);
		\vertex [right = 0.75 of a] (v1);
		\vertex [above right = of v1] (v2);
		\vertex [below right = of v1] (v3);
		\vertex [right = 0.75 of v2] (p1);
		\vertex [right = 0.75 of v3] (p2);
		\diagram*{
		(a) -- [scalar] (v1) -- [fermion] (v2) -- [fermion] (v3) -- [fermion] (v1),
		(v2) -- [photon] (p2),
		(v3) -- [photon] (p1),
		};
	\end{feynman}
	\end{tikzpicture}.
\end{align}
Its evaluation is similar to the chiral anomaly computation, and we obtain
\begin{align}
	i\mathcal{M} &= \frac{2i\alpha}{\pi} \epsilon^{\alpha\beta\mu\nu}p_{1\mu}p_{2\nu}\epsilon_\alpha^*(p_1) \epsilon_\beta^*(p_2)
	\nonumber \\
	&\times
	\int_0^1dx\int_0^{1-x}dy\left[1-\frac{m_\ell^2}{m_\ell^2 - xy m_a^2}\right].
\end{align}
This expression reproduces the chiral anomaly in the limit $m_\ell \ll m_a$,
while it vanishes in the opposite limit $m_\ell \gg m_a$.
In general, the anomaly equation for a massive fermion contains two contributions; 
$\partial_\mu (\bar{\ell} \gamma^\mu\gamma_5 \ell)$ and $m_\ell \bar{\ell}i\gamma_5 \ell$.
In the high energy (or massless) limit, the anomaly equation, \emph{i.e.}, the $F\tilde{F}$ operator, is saturated by the former contribution, while
in the low energy (or heavy mass) limit, it is saturated by the latter contribution.
One can perform the Feynman parameter integrals  analytically, leading to~\cref{eq:a2photon} after the phase space integration.

\bibliographystyle{utphys}
\bibliography{ref}

\end{document}